\documentclass[aps,prx,twocolumn,superscriptaddress]{revtex4-2}
\usepackage{graphicx,amssymb,amsmath,amsfonts,empheq}
\usepackage[dvipsnames]{xcolor}
\usepackage{comment}
\usepackage{color}
\usepackage{braket}
\usepackage{latexsym}
\usepackage{upgreek}
\usepackage{dsfont}
\usepackage{bm}
\usepackage{makecell}
\usepackage{multirow}
\usepackage{enumitem}
\usepackage[normalem]{ulem}
\usepackage{url}
\usepackage{soul}
\usepackage{hyperref}
\hypersetup{
colorlinks = true,
linkcolor = [rgb]{0,0,0}, 
citecolor = [rgb]{0.13,0.55,0.13},
urlcolor = [rgb]{0.25, 0.41, 0.88}}
\usepackage{bbm}

\usepackage{tikz}
\usepackage{tikz-cd}
\usetikzlibrary{arrows}
\usetikzlibrary{intersections}
\usetikzlibrary{shapes.geometric}
\usetikzlibrary{decorations.pathmorphing, patterns,shapes}
\usetikzlibrary{decorations.markings}

\tikzset{
	mid arrow/.style={postaction={decorate,decoration={
				markings,
				mark=at position .575 with {\arrow[#1]{stealth}}
	}}},
	near arrow/.style={postaction={decorate,decoration={
				markings,
				mark=at position .275 with {\arrow[#1]{stealth}}
	}}},
	far arrow/.style={postaction={decorate,decoration={
				markings,
				mark=at position .800 with {\arrow[#1]{stealth}}
	}}},
}

\tikzset{meter/.append style={draw, inner sep=6.5, rectangle, font=\vphantom{A}, minimum width=25, line width=.8,
 path picture={
 \draw[black] ([shift={(.1,.15)}]path picture bounding box.south west) to[bend left=50] ([shift={(-.1,.15)}]path picture bounding box.south east);
 \draw[black,-latex] ([shift={(0,.1)}]path picture bounding box.south) -- ([shift={(.3,-.1)}]path picture bounding box.north);}}}

\renewcommand{\leq}{\leqslant}
\renewcommand{\geq}{\geqslant}

\newcommand{\tr}{\mathrm{tr}}
\newcommand{\sgn}{\operatorname{sgn}}

\newcommand{\bbZ}{\mathbb{Z}}

\newcommand{\calC}{\mathcal{C}}

\newcommand{\calE}{\mathcal{E}}
\newcommand{\calF}{\mathcal{F}}

\newcommand{\calN}{\mathcal{N}}

\newcommand{\calP}{\mathcal{P}}

\newcommand{\calS}{\mathcal{S}}

\newcommand{\calZ}{\mathcal{Z}}

\newcommand{\eqnref}[1]{Eq.~\eqref{#1}}
\newcommand{\figref}[1]{Fig.~\ref{#1}}
\newcommand{\appref}[1]{Appendix~\ref{#1}}
\newcommand{\tabref}[1]{Table~\ref{#1}}
\newcommand{\secref}[1]{Sec.~\ref{#1}}

\begin{document}

\title{Diagnostics of mixed-state topological order and breakdown of quantum memory}

\author{Ruihua Fan}
\thanks{RF and YB contributed equally to this work.}
\affiliation{Department of Physics, Harvard University, Cambridge, MA 02138, USA}
\author{Yimu Bao}
\thanks{RF and YB contributed equally to this work.}
\affiliation{Department of Physics, University of California, Berkeley, CA 94720, USA}
\author{Ehud Altman}
\affiliation{Department of Physics, University of California, Berkeley, CA 94720, USA}
\affiliation{Materials Sciences Division, Lawrence Berkeley National Laboratory, Berkeley, CA 94720, USA}
\author{Ashvin Vishwanath}
\affiliation{Department of Physics, Harvard University, Cambridge, MA 02138, USA}

\begin{abstract}

Topological quantum memory can protect information against local errors up to finite error thresholds. Such thresholds are usually determined based on the success of decoding algorithms rather than the intrinsic properties of the mixed states describing corrupted memories. Here we provide an intrinsic characterization of the breakdown of topological quantum memory, which both gives a bound on the performance of decoding algorithms and provides examples of topologically distinct mixed states. We employ three information-theoretical quantities that can be regarded as generalizations of the diagnostics of ground-state topological order, and serve as a definition for topological order in error-corrupted mixed states. We consider the topological contribution to entanglement negativity and two other metrics based on
quantum relative entropy and coherent information. In the concrete example of the 2D Toric code with local bit-flip and phase errors, we map three quantities to observables in 2D classical spin models and analytically show they all undergo a transition at the same error threshold. This threshold is an upper bound on that achieved in any decoding algorithm and is indeed saturated by that in the optimal decoding algorithm for the Toric code.

\end{abstract}

\maketitle

\section{Introduction}

The major roadblock to realizing quantum computers is the presence of errors and decoherence from the environment which can only be overcome by adopting quantum error correction (QEC) and fault tolerance~\cite{Gottesman2009}. A first step would be the realization of robust quantum memories~\cite{Calderbank:1995dw,Steane:1996ghp,Terhal_2015}.
Topologically ordered systems in two spatial dimensions, owing to their long-range entanglement and consequent degenerate ground states, serve as a promising candidate~\cite{ChetanReview,xiaogangReview,Kitaev:1997wr,Fujii:2015wia}.
A paradigmatic example is the surface code~\cite{Bravyi:1998sy,Dennis:2001nw}, whose promise as a robust quantum memory has stimulated recent interest in its realization in near-term quantum simulators~\cite{Nigg_2014, Satzinger:2021eqy,Verresen:2020dmk,Semeghini:2021wls, Bluvstein:2021jsq,GoogleQuantumAI:2022fyn,Andersen:2022xmz}.

One of the central quests is to analyze the performance of topological quantum memory under \emph{local} decoherence.
In the case of surface code and other topological codes with local errors, it has been shown that the stored information can be decoded reliably up to a finite error threshold~\cite{Dennis:2001nw,Wang:2002ph,Katzgraber:2009zz,Bombin:2012jk,Kubica:2018rab,Flammia2021}. 
Namely, as the error rate increases, the success probability of the decoding algorithm drops to zero at a critical value, which depends on the choice of the algorithm.
It is then natural to ask whether these decoding transitions stem from an intrinsic error-induced singularity in the mixed states.
If so, how to probe this intrinsic transition?

The intrinsic characterization has at least two important consequences. 
First, the critical error rate for the intrinsic transition should furnish an upper bound for decoding algorithms.
The algorithmic dependence of the decoding thresholds is a mere reflection of the suboptimality of specific algorithms.
Second, the correspondence between successful decoding and intrinsic  properties of the quantum state acted upon by errors points to the existence of topologically distinct mixed states.
In another word, answering this question amounts to relating the breakdown of topological quantum memory to a transition in the mixed-state topological order.
Progress towards this goal lies in quantifying the residual long-range entanglement in the error-corrupted mixed state. We will consider quantities that are motivated from both perspectives and explore their unison.

In this work, we investigate three information-theoretical diagnostics: (i) quantum relative entropy between the error-corrupted ground state and excited state; (ii) coherent information; (iii) topological entanglement negativity.
The first two are natural from the perspective of quantum error correction (QEC). More specifically, the quantum relative entropy quantifies whether errors ruin orthogonality between states~\cite{kitaev2002classical}, and coherent information is known to give robust necessary and sufficient conditions on successful QEC~\cite{Schumacher:1996dy,Schumacher:2001,Horodecki:2005ehk}.
The third one is a basis-independent characterization of long-range entanglement in mixed states and is more natural from the perspective of mixed-state topological order.
This quantity has been proposed to diagnose topological orders in Gibbs states~\cite{Lu:2019owx,Lu:2022yad}, which changes discontinuously at the critical temperature. We borrow and apply this proposal to error-corrupted states. Our transition occurs in two spatial dimensions at a finite error rate, in contrast to the finite temperature transitions in four spatial dimensions.

The presence of three seemingly different diagnostics raises the question of whether they all agree and share the same critical error rate.
Satisfyingly, we indeed find this to be the case in a concrete example, surface code with bit-flip and phase errors.
The $n$-th R\'enyi version of the three quantities can be formulated in a \emph{classical} two-dimensional statistical mechanical model of $(n-1)$-flavor Ising spins, which exhibits a transition from a paramagnetic to a ferromagnetic phase as the error rate increases.
The three quantities are mapped to different probes of the ferromagnetic order and must undergo the transition simultaneously, which establishes their consistency in this concrete example.

Interestingly, the statistical mechanical model derived for the information-theoretic diagnostics is exactly dual to the random-bond Ising model (RBIM) that governs the decoding transition of the algorithm proposed in \cite{Dennis:2001nw}.
This duality implies that the error threshold of the algorithm in \cite{Dennis:2001nw} saturates the upper bound.
Therefore, it confirms that this decoding algorithm is optimal, and its threshold reflects the intrinsic properties of the corrupted state.
We remark that mappings to statistical mechanical models have been tied to obtaining error thresholds of decoding algorithms~\cite{Dennis:2001nw,Wang:2002ph,Katzgraber:2009zz,Bombin:2012jk,Kubica:2018rab,Flammia2021}. Here such mappings arise from characterizing intrinsic properties of the error corrupted mixed state.

The rest of the paper is organized as follows. 
\secref{sec:diagnostics} gives a concrete definition of the error-corrupted states and introduces the three diagnostics. 
\secref{sec:example} studies the concrete example, the 2D Toric code subject to local bit-flip and phase errors.
We close with discussions in \secref{sec:discussion}.

\section{Setup and diagnostics}
\label{sec:diagnostics}

In this section, we begin with introducing the error-corrupted mixed state.
We show that any operator expectation value in a single-copy corrupted density matrix cannot probe the transition, and instead one needs to consider the non-linear functions of the density matrix.
Next, we introduce three information-theoretic diagnostics of the transition: (i) quantum relative entropy; (ii) coherent information; (iii) topological entanglement negativity.
These quantities generalize the diagnostics of ground-state topological order.

\subsection{Error-corrupted mixed state}
\label{sec:error corrupted state}

The type of mixed state we consider in the paper describes a topologically ordered ground state $\ket{\Psi_0}\bra{\Psi_0}$ subject to local errors 
\begin{equation}
    \rho = \calN[\ket{\Psi_0}\bra{\Psi_0}] = \prod_i \calN_{i} [\ket{\Psi_0}\bra{\Psi_0}]\,,
\end{equation}
where the quantum channel $\calN_i$ models the local error at site $i$ and is controlled by the error rate $p$.
We refer to $\rho$ as the error-corrupted mixed state.

\begin{figure}[t]
\centering
\includegraphics[width=0.46\textwidth]{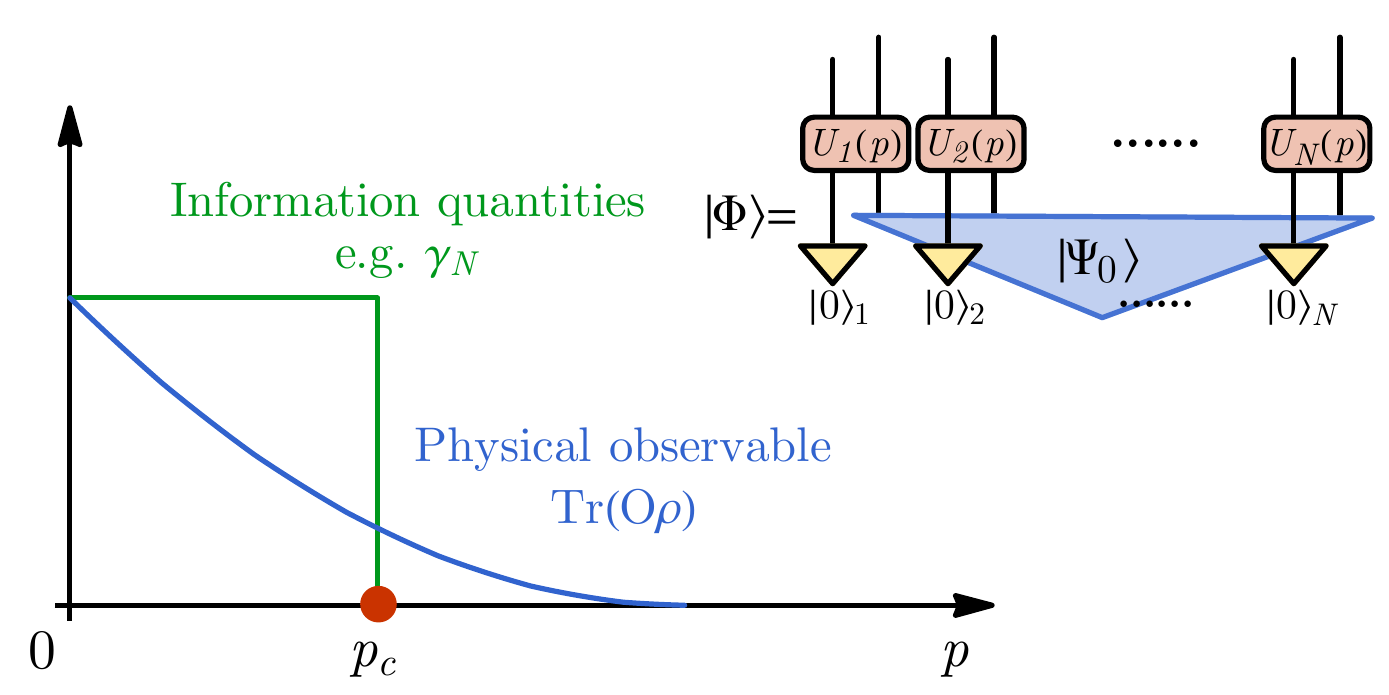}
\caption{Physical observables verses information quantities in error corrupted states. Each error corrupted state can be obtained from applying local unitaries to the system (topological order) plus ancilla qubits (trivial product state). Thus, physical observables must be smooth functions of the error rate $p$. In contrast, information quantities, e.g. the topological entanglement negativity $\gamma_N$, can have discontinuities that identify the many-body singularities.}
\label{fig:intro}
\end{figure}

The transition in the corrupted state, if exists, cannot be probed by the operator expectation value in a single-copy density matrix.
To demonstrate it, we purify the corrupted state by introducing one ancilla qubit prepared in $\ket{0}_i$ for each physical qubit at site $i$.
The physical and ancilla qubits are coupled locally via unitary $U_i(p)$ such that tracing out the ancilla qubits reproduces the corrupted state $\rho$.
This leads to a purification
\begin{align}
    \ket{\Phi} = \prod_i U_i(p) \ket{\Psi_0} \left(\otimes_i \ket{0}_i\right),\label{eq:naive_purification}
\end{align}
which is related to the topologically ordered state by a depth-1 unitary circuit on the extended system [see Fig.~\ref{fig:intro}].
It follows that the expectation value of \emph{any} operator supported on a large but finite region of the physical qubits, e.g., a Wilson loop operator, must be a smooth function of the error rate [see Fig.~\ref{fig:intro} for a schematics].
Thus, it is indispensable to consider the non-linear functions of the density matrix, e.g. quantum information quantities, to probe the transition in the corrupted state.
This property holds when $\rho$ describes a general mixed state in the ground-state subspace under local errors.

We remark that the above argument does not prevent observables in a single-copy density matrix from detecting topological order in finite-temperature Gibbs states~\cite{Hasting2011FiniteT}.
The key difference is the purifications of the Gibbs states at different temperatures are not necessarily related by finite-depth circuits.

\subsection{Quantum relative entropy}\label{sec:diagnostics relative entropy}

Anyon excitations are crucial for storing and manipulating quantum information in a topologically ordered state. For example, to change the logical state of the code one creates a pair of anyons out of the vacuum and separates them to opposite boundaries of the system. The first diagnostic tests if the process of creating a pair of anyons and separating them by a large distance gives rise to a distinct state in the presence of decoherence.

Specifically, we want to test if the corrupted state $\rho=\mathcal{N}[\rho_0]$ is sharply distinct from $\rho_\alpha=\mathcal{N}[w_\alpha(\calP)\rho_0 w_\alpha(\calP)^\dagger]$ for $\rho_0$ in the ground state subspace. In the second state, $w_\alpha(\calP)$ is an open string operator that creates an anyon $\alpha$ and its anti-particle $\alpha'$ at the opposite ends of the path $\calP$. We use the \emph{quantum relative entropy} as a measure for the distinguishability of the two states
\begin{align}
    D(\rho||\rho_\alpha) := \tr \rho \log \rho - \tr \rho \log \rho_\alpha\,.
\end{align}
In absence of errors the relative entropy is infinite because the two states are orthogonal, and it decreases monotonically with the error rate~\cite{Lieb:1973cp,Araki:1976zv,Lindblad1975}.
Below the critical error rate, however, the states should remain perfectly distinguishable if the anyons are separated by a long distance. Therefore we expect the relative entropy to diverge as the distance between the anyons is taken to infinity. 
Above the critical error rate on the other hand we expect the relative entropy to saturate to a finite value reflecting the inability to perfectly distinguish between the two corrupted states.
In this regard, the relative entropy describes whether anyon excitations remain well-defined and is a generalization of the Fredenhagen-Marcu order parameter for ground state topological order~\cite{Fredenhagen:1983sn,Fredenhagen:1985ft,Fredenhagen:1987gi,Gregor:2010ym}.

To facilitate calculations, we consider a specific sequence of the R\'enyi relative entropies 
\begin{align}
    D^{(n)}(\rho||\rho_\alpha) := \frac{1}{1-n} \log \frac{\tr \rho\rho_\alpha^{n-1}}{\tr \rho^n}, \label{eq:renyi_rel}
\end{align}
which recovers $D(\rho||\rho_\alpha)$ in the limit $n\rightarrow 1$. 
In Sec.~\ref{sec:example} we map the relative entropies $D^{(n)}$ in the corrupted Toric code to order parameter correlation functions in an effective statistical mechanical model, which is shown to exhibit the expected behavior on two sides of the critical error rate.

\subsection{Coherent information}
\label{sec:diagnostics coherent info}

The basis for protecting quantum information in topologically ordered states is encoding it in the degenerate ground state subspace. 
The second diagnostic we consider is designed to test the integrity of this protected quantum memory.

We use the \emph{coherent information}, as a standard metric for the amount of recoverable quantum information after a decoherence quantum channel~\cite{Schumacher:1996dy,Schumacher:2001,Horodecki:2005ehk}.
In our case, the relevant quantum channel consists of the following ingredients illustrated below: (i) a unitary operator $U$ that encodes the state of the logical qubits in the input $R$ into the ground state subspace; (ii) a unitary coupling $U_{QE}$ of the physical qubits $Q$ to environment qubits $E$, which models the decoherence.
The coherent information in this setup is defined as
\begin{equation}
    I_c(R\rangle Q) := S_Q - S_{QR}.\quad
    \begin{tikzpicture}[scale=0.6,baseline={(current bounding box.center)}]
    	\draw[thick] (-1, 2)node[left]{\scriptsize$R$} -- (-1, 0) -- (0.3, 0) -- (0.3,1);
    	\draw[thick] (0.8,1) -- (0.8,2)node[left]{\scriptsize$Q$};
    	\draw[thick] (1.5,0) -- (1.5,1);
    	\draw[thick] (2.5,0) -- (2.5, 2)node[right]{\scriptsize$E$};
    	\draw[dashed] (0.8,1.6) -- node[above] {\scriptsize$U_{QE}$} ++ (1.7,0);
    	\filldraw [fill=white, thick, rounded corners=0.2cm] (0, 0.5) rectangle ++(1.8, 0.8) node [midway] {\scriptsize$U$};
    \end{tikzpicture}
\end{equation}
Here $S_Q$ and $S_{RQ}$ are the von Neumann entropies of the systems $Q$ and $RQ$ respectively, and we use the Choi map to treat the input $R$ as a reference qubit in the output.
It follows from subadditivity that the coherent information is bounded by the amount of encoded information in the degenerate ground state subspace, i.e. $-S_R \leq I_c \leq S_R$.
In the absence of errors, $I_c = S_R$, and we expect this value to persist as long as the error rate is below the critical value. 
Above the critical error rate, we expect $I_c< S_R$, indicating the loss of encoded information. 
We remark that the recoverable information is also used to characterize the robustness of quantum memory based on the edge mode in 1D Kitaev chain~\cite{Mazza:2013qna}.

Physically the coherent information is closely related and expected to undergo a transition at the same point as the relative entropy discussed above. The quantum information is encoded by separating anyon pairs across the system. It stands to reason that if this state remains perfectly distinguishable from the original state, as quantified by the relative entropy, then the quantum information encoded in this process is preserved.

The critical error rate for preserving the coherent information is an upper bound for the threshold of any QEC algorithms
\begin{equation}
    p_{c} \geq p_{c,\text{algorithm}}\,.\label{eq:pc_upper}
\end{equation}
The key point is that coherent information is non-increasing upon quantum information processing and cannot be restored once it is lost.
Thus, a successful QEC requires $I_c = S_R$.
Moreover, the QEC algorithm involves syndrome measurements that are non-unitary and generically do not access the full coherent information in the system giving rise to a lower error threshold. 

To facilitate calculations and mappings to a statistical mechanical model we will need the R\'enyi coherent information
\begin{align}
    I_c^{(n)} := S^{(n)}_Q - S^{(n)}_{RQ} = \frac{1}{n-1} \log \frac{\tr \rho_{RQ}^n}{\tr \rho_{Q}^n}, \label{eq:renyi_Ic}
\end{align}
which approaches $I_c$ in the limit $n \to 1$.
In the example of Toric code with incoherent errors discussed in Sec.~\ref{sec:example}, we show that $I_c^{(n)}$ takes distinct values in different phases.

\subsection{Topological entanglement negativity}
\label{sec:diagnostics negativity}

The topological entanglement entropy provides an intrinsic bulk probe of ground state topological order and does not require a priori knowledge of the anyon excitations. The third diagnostic we consider generalizes this notion to the error-corrupted mixed state.

A natural quantity often used to quantify entanglement in mixed states, is the logarithmic negativity of a sub-region $A$~\cite{Peres:1996dw,Horodecki:1996nc,Vidal:2002zz}
\begin{equation}
	\calE_{A}(\rho) := \log ||\rho^{T_A} ||_1,
\end{equation}
where $\rho^{T_A}$ is the partial transpose on the subsystem $A$ and $\Vert\cdot\Vert_1$ denotes the trace ($L_1$) norm.
The logarithmic negativity coincides with the R\'enyi-1/2 entanglement entropy for the pure state and is non-increasing with the error rate of the channel, a requirement that any measure of entanglement must satisfy~\cite{nielsen2002quantum,Plenio:2005cwa}.
The logarithmic negativity was previously used in the study of ground state topological phases~\cite{Wen:2016snr,Wen:2016bla,Shapourian:2016cqu} and more recently for detecting  topological order in finite temperature Gibbs states~\cite{Lu:2019owx,Lu:2022yad}.

We expect that the universal topological contribution to the entanglement ~\cite{Kitaev:2005dm,LevinWen2005} will survive in the corrupted mixed state below a critical error rate and can be captured by the logarithmic negativity. Thus, the conjectured form of this quantity is
\begin{align}
	\calE_{A} = c|\partial A| - \gamma_N + \ldots,
\end{align}
where $|\partial A|$ is the circumference of the region $A$, $c$ is a non-universal coefficient, and ellipsis denotes terms that vanish in the limit $|\partial A| \rightarrow \infty$. 
The constant term $\gamma_N$ is the \emph{topological entanglement negativity} of a simply connected subregion. 
It is argued to be a topological invariant that cannot come from local contributions to the entanglement due to the conversion property $\mathcal{E}_{A} = \mathcal{E}_{\bar{A}}$, i.e. negativity of a subsystem is equal to that of the complement~\cite{Grover:2011fa,Lu:2019owx}. 
Let us repeat the argument here for the reader's convenience. 
We assume that the non-topological part, arising from local contributions, can be written as an integral along the entanglement cut, $\calE_{A,local} = \int_{\partial A} f(\kappa, \partial\kappa) dl$, where $f(\kappa, \partial\kappa)$ depends on the extrinsic curvature $\kappa$ of the cut.
For a smooth and large entanglement cut, one can perform a Taylor expansion $f(\kappa, \partial \kappa) = f_0 + f_1\kappa + \ldots$, which integrates to $c|\partial A| + c_1 + c_2 |\partial A|^{-1} + \ldots$.
Notably, the extrinsic curvature changes its sign when transforming $A$ to $\bar{A}$, necessitating the vanishing of all odd-order terms to ensure $\mathcal{E}_{A} = \mathcal{E}_{\bar{A}}$. 
In particular, $f_1 = 0$ and $c_1 = 0$.
Thus, local contributions cannot produce a constant term in the negativity.
In contrast, the von Neumann entropy of a subregion in the error-corrupted mixed state exhibits a volume-law scaling, and its constant piece is not topological because $S_A \neq S_{\bar{A}}$.

To facilitate the calculation of the negativity, we consider the R\'enyi negativity of even order
\begin{align}
    \calE^{(2n)}_{A}(\rho) := \frac{1}{2-2n} \log \frac{\tr (\rho^{T_A})^{2n}}{\tr \rho^{2n}}\,.\label{eq:renyi_neg}
\end{align}
The logarithmic negativity is recovered in the limit $2n \to 1$.
Here, we choose a particular definition of the R\'enyi negativity such that it exhibits an area-law scaling in the corrupted state.
In Sec.~\ref{sec:example}, we show explicitly that in the Toric code the topological part $\gamma_N^{(2n)}$  of the R\'enyi negativity takes a quantized value $\log 2$ in the phase where the quantum memory is retained and vanishes otherwise.

To summarize, we expect the topological negativity takes the same universal value as the topological entanglement entropy in the uncorrupted ground state and drops sharply to a lower value at a critical error rate.
It is \textit{a priori} not clear, however, that the transition in the negativity must occur at the same threshold as that marks the transition of the other two diagnostics we discussed. 
In Sec.~\ref{sec:example} we show, through mapping to a statistical mechanical model that, in the example of the Toric code, a single phase transition governs the behavior of all three diagnostics.

\begin{table*}[th!]
\centering
\begin{tabular}{|c|c|c|c|}
\hline
\hline
Diagnostics & Observable & PM & FM \\
\hline
$D^{(n)}$ & \makecell{Logarithm of \\
order parameter correlation function}
& $ O (|i_l - i_r|)$ & $O(1)$ \\
\hline
$I_c^{(n)}$ & 
\makecell{Related to the excess free energy for \\
domain walls along non-contractible loops}
& $2\log 2$ & $0$ \\
\hline
$\mathcal{E}_{A}^{(2n)}$ & 
\makecell{Excess free energy for \\ aligning spins on the boundary of $A$}
& $c|\partial A|/\xi - \log 2$ & $c|\partial A|/\xi$ \\
\hline
\hline
\end{tabular}
\caption{Dictionary of the mapping.
The R\'enyi-$n$ version of the diagnostics of topological order in error corrupted states and their corresponding observables in $(n-1)$-flavor Ising models are listed in the first and second columns, respectively.
We consider 2D Toric code subject to one type of incoherent error (bit-flip or phase errors).
The asymptotic behaviors of these diagnostics in the paramagnetic (PM) and ferromagnetic (FM) phases of the spin model are provided.
}
\label{tab:dict}
\end{table*}

\section{Example: Toric code under bit-flip and phase errors}
\label{sec:example}

In this section, we use the three information-theoretical diagnostics to probe the distinct error-induced phases in the 2D Toric code under bit-flip and phase errors.
In particular, we develop 2D classical statistical mechanical models to analytically study the R\'enyi-$n$ version of the diagnostics in this example.
The statistical mechanical models involve $(n-1)$-flavor Ising spins and undergo ferromagnetic phase transitions as a function of error rates.
We show that the three diagnostics map to distinct observables that all detect the ferromagnetic order and undergo the transition simultaneously.
We remark that our results also apply to the planar surface code.

In Sec.~\ref{sec:tc+error}, we introduce the Toric code and the error models.
We derive the statistical mechanical models in Sec.~\ref{sec:stat-mech model} and analyze the phase transition in Sec.~\ref{sec:phase transitions}.
Sec.~\ref{sec:stat-mech diagnostics} discusses the three diagnostics and their corresponding observables in the statistical mechanical models. See \tabref{tab:dict} for a summary.
We discuss the replica limit $n \to 1$ in Sec.~\ref{sec:duality}.

\subsection{Toric code and error model}\label{sec:tc+error}

We consider the 2D Toric code on an $L\times L$ square lattice with periodic boundary conditions.
This code involves $N = 2L^2$ physical qubits on the edges of the lattice, and its code space is given by the ground state subspace of the Hamiltonian
\begin{equation}
    H_{\text{TC}} = - \sum_s A_s - \sum_p B_p\,,
\end{equation}
where $A_s$ and $B_p$ are mutually commuting operators associated with vertices and plaquettes
\begin{equation}
    A_s = \prod_{\ell \in \text{star}(s)} X_\ell\,, \quad B_p = \prod_{\ell \in \text{boundary}(p)} Z_\ell\,.
\end{equation}
Here, $X_\ell$ and $Z_\ell$ denote the Pauli-X and Z operators on edge $\ell$, respectively.
The ground state satisfying $A_s\ket{\Psi} = B_p \ket{\Psi} = \ket{\Psi}$ is four-fold degenerate and can encode two logical qubits.

We consider specific error channels describing uncorrelated single-qubit bit-flip and phase errors
\begin{equation}
\begin{aligned}
    \mathcal{N}_{X,i}[\rho] &= (1 - p_x)\rho + p_x X_i \rho X_i\,, \\
    \mathcal{N}_{Z,i}[\rho] &= (1 - p_z)\rho + p_z Z_i \rho Z_i\,,
\end{aligned}\label{eq:1qbit-error_channel}
\end{equation}
where the Pauli-$X$ ($Z$) operator acting on the Toric code ground state creates a pair of $m$ ($e$) anyons on the adjacent plaquettes (vertices), $p_x$ and $p_z$ are the corresponding error rates.
The corrupted state reads
\begin{equation*}
	\rho = \mathcal{N}_{X} \circ \mathcal{N}_{Z} [\rho_0]\,,
\end{equation*}
where $\calN_{X(Z)} = \prod_i \calN_{X(Z),i}$.
We assume that the error rate is uniform throughout our discussion.
We remark that the error channels in Eq.~\eqref{eq:1qbit-error_channel} do not create coherent superposition between states with different anyon configurations and are referred to as incoherent errors. Pauli-$Y$ errors create anyons incoherently and can also be analyzed using our framework.

\subsection{Statistical mechanical models}
\label{sec:stat-mech model}

Here, we map the $n$-th moment of the corrupted density matrix $\tr \rho^n$ to the partition function of the $(n-1)$-flavor Ising model.
In this statistical mechanical model, one can analyze the singularity in the R\'enyi version of the three diagnostics, which will be presented in Sec.~\ref{sec:stat-mech diagnostics}.

To begin, we consider the maximally mixed state in the ground state subspace
\begin{equation}
	\rho_0 = \frac{1}{4}\prod_s \frac{1 + A_s}{2} \prod_p \frac{1 + B_p}{2}\,.
\end{equation}
We note that the choice of the ground state $\rho_0$ determines the boundary conditions in the resulting model and does not affect the location of the critical point.
For our purpose here, it is convenient to write $\rho_0$ in a loop picture
\begin{equation}
	\begin{tikzpicture}[scale=0.75, baseline={(current bounding box.center)}]
		\definecolor{myred}{RGB}{240,83,90};
		\definecolor{myblue}{RGB}{93,123,189};
		\small
		\foreach \y in {0,0.5,...,1.5,2}{
			\draw[black!60] (-0.2,\y) -- (2.2,\y);
		}
		\foreach \x in {0,0.5,...,2}{
			\draw[black!60] (\x,-0.2) -- (\x,2.2);
		}
		\draw[myred,line width=1] (0,0) -- (0,1) -- (1,1) -- (1,0.5) -- (0.5,0.5) -- (0.5,0) -- (0,0);
		\draw[myred,line width=1] (0.5,2) -- (1.5,2) -- (1.5,1.5) -- (0.5,1.5) -- (0.5,2);
		\draw[myblue,line width=1.5, dashed] (0.75,0.25) -- (1.75,0.25) -- (1.75,1.25) -- (0.75,1.25) -- (0.75,0.25);
		\node[myred] at (-0.27,0.75) {$g_z$};
		\node[myblue] at (2.25,0.25) {$g_x$};
	\end{tikzpicture}\qquad
	\rho_0 = \frac{1}{2^N} \sum_{g_z} g_z \sum_{g_x} g_x\,,\label{eq:rho0_loop}
\end{equation}
where $g_z$ and $g_{x}$ are $Z$ and $X$ loops on the original and dual lattice given by the product of $A_s$ and $B_p$ operators, respectively.
The summation runs over all possible loop configurations. 
In what follows, we will use $g_{x(z)}$ to denote both the operators and the loop configurations. The meaning will be clear in the context.

Two error channels act on the loop operators $g_x,g_z$ by only assigning a real positive weight:
\begin{equation*}
	\begin{aligned}
		\mathcal{N}_{X,i}[g_z] &= \left\{\begin{array}{rc}
			(1 - 2p_x) g_z & Z_i \in g_z \\
			g_z & Z_i \notin g_z
		\end{array}\right. , \\
		\mathcal{N}_{Z,i}[g_x] &= \left\{\begin{array}{rc}
			(1 - 2p_z) g_x &  X_i \in g_x \\
			g_x & X_i \notin g_x
		\end{array}\right. .
	\end{aligned}
\end{equation*}
Thus, the corrupted state remains a superposition of loop operators
\begin{equation}
	\rho 
	= \frac{1}{2^N}\sum_{g_x, g_z} e^{-\mu_x|g_x|-\mu_z|g_z|}g_xg_z,\label{eq:rho_loop}
\end{equation}
where $|g_{x(z)}|$ denotes the length of the loop, and $\mu_{x(z)} = -\log(1 - 2p_{z(x)})$ can be understood as the line tension. 
Using \eqnref{eq:rho_loop}, it is straightforward to see that the expectation values of operators, such as the Wilson loop and open string, behave smoothly as the error rate increases, in consistence with the general argument in \secref{sec:error corrupted state}.

Using this loop picture \eqnref{eq:rho_loop}, we can write the $n$-th moment as
\begin{equation}
\begin{aligned}
    \tr \rho^n = \frac{1}{2^{nN}}\sum_{\{g_{x}^{(s)},g_z^{(s)}\}} &\tr \Big(\prod_{s=1}^n g_{x}^{(s)} g_{z}^{(s)}\Big) \\
    &e^{\sum_s -\mu_x|g_{x}^{(s)}|-\mu_z|g_z^{(s)}|},
    \label{eq:nth_moment_loop}
\end{aligned}
\end{equation}
where $g_{x(z)}^{(s)}$, $s = 1,2,\cdots,n$ is the $X(Z)$ loop operator from the $s$-th copy of density matrix.
The product of loop operators in Eq.~\eqref{eq:nth_moment_loop} has a nonvanishing trace only if the products of $X$ and $Z$ loops are proportional to identity individually, which leads to two independent constraints 
\begin{equation}
	\label{eq:constraint on loops}
	g_{a}^{(n)} = \prod_{s = 1}^{n-1} g_{a}^{(s)}\,,\quad a = x,z\,.
\end{equation}

The $n$-th moment factorizes into a product of two partition functions
\begin{equation}
\label{eq:factorization}
	\tr\rho^n = \frac{1}{2^{(n-1)N}} \mathcal{Z}_{n,x}\mathcal{Z}_{n,z}\,,
\end{equation}
where $\mathcal{Z}_{n,a} = \sum_{\{g^{(s)}_{a}\}} e^{-H_{n,a}}$ with $a = x,z$ is a statistical mechanical model that describes fluctuating $X(Z)$ loops with a line tension.
The Hamiltonian takes the form
\begin{equation}
	H_{n,a} = \mu_{a}\Big(\sum_{s=1}^{n-1} \big|g_a^{(s)}\big| + \big|\prod_{s=1}^{n-1} g_{a}^{(s)}\big|\Big)\,.
\label{eq:stat-mech_spin}
\end{equation}
Here, we have imposed the constraints \eqref{eq:constraint on loops}, and the summation in each partition function runs over the loop configurations only in the first $n-1$ copies.

The loop model can be mapped to a statistical mechanical model of $n-1$ flavors of Ising spins with nearest neighbor ferromagnetic interactions.
The mapping is established by identifying the loop configuration $g_a^{(s)}$ with $s = 1,2,\ldots,n-1$ with domain walls of Ising spins.
Specifically, for a $Z$ loop configuration on the original lattice, we associate a Ising spin configuration $\sigma_i$ on the dual lattice such that 
$$
\begin{tikzpicture}[scale=0.75, baseline={(current bounding box.center)}]
\definecolor{myred}{RGB}{240,83,90};
\definecolor{myblue}{RGB}{93,123,189};
\small
\foreach \y in {0,0.7,1.4,2.1}{
	\draw[black!35] (-0.2,\y) -- (2.2,\y);
}
\foreach \x in {0,0.7,1.4,2.1}{
	\draw[black!35] (\x,-0.2) -- (\x,2.2);
}
\draw[myblue,line width=0.6] (-0.2,0.7) -- (0.7,0.7) --  (0.7,1.4) -- (2.1,1.4) -- (2.1,2.2);
\draw[myblue,line width=1.5] (0.7,1.4) -- (1.4,1.4);
\filldraw (1.05,1.05) circle (0.03) node[right]{\scriptsize$\sigma_i^{(s)}$};
\filldraw (1.05,1.8) circle (0.03) node[right]{\scriptsize$\sigma_j^{(s)}$};
\draw[dotted,line width=1] (1.05,1.05) -- (1.05,1.8);
\node[myblue] at (0.3,1.8) {$g_{z,\ell}^{(s)}$};
\end{tikzpicture}
\qquad
\left|g_{z,\ell}^{(s)}\right| = \left(1 - \sigma_{i}^{(s)}\sigma_j^{(s)}\right)/2\,,
$$ 
where $i,j$ are connected by the link dual to $\ell$, and $|g_{z,\ell}^{(s)}|$ is a binary function that counts the support of loop on link $\ell$.
The total length of the loop is given by $|g_z^{(s)}| = \sum_\ell |g_{z,\ell}^{(s)}|$.
Similarly, we can define the Ising spins on the original lattice that describe the $X$ loop configuration on the dual lattice.

In terms of the Ising spins, the effective Hamiltonian is given by
\begin{equation}
    H_{n,a} = -J_a\sum_{\langle i,j\rangle} \left(\sum_{s = 1}^{n-1} \sigma_i^{(s)}\sigma_j^{(s)} + \prod_{s = 1}^{n-1} \sigma_i^{(s)} \sigma_j^{(s)} \right)\label{eq:Heff_loop}
\end{equation}
with a ferromagnetic coupling
$J_{x(z)} = -\log\sqrt{1-2p_{z(x)}}\,.$
In what follows, we refer to this model as the \emph{$(n-1)$-flavor Ising model}.
We remark that the model exhibits a global symmetry $G^{(n)} = (\mathbb{Z}_2^{\otimes n} \rtimes \mathcal{S}_n)/\mathbb{Z}_2$, where $\mathcal{S}_n$ is the permutation symmetry over $n$ elements.
As is shown below, increasing the error rate the model undergoes a paramagnetic-to-ferromagnetic transition that completely breaks the $G^{(n)}$ symmetry.

\subsection{Phase transitions}\label{sec:phase transitions}

\begin{figure}
    \centering
    \includegraphics[width=0.46\textwidth]{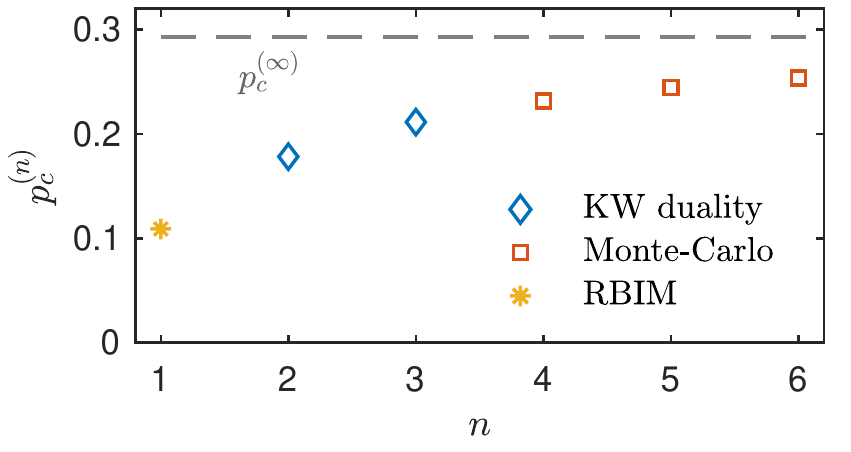}
    \caption{Critical error rates for various R\'enyi index $n$. $p_c^{(2)} \approx 0.178$ and $p_c^{(3)} \approx 0.211$ are determined by the exact solution (blue diamonds). For $n \geq 4$, $p_c^{(n)}$ is determined by calculating the crossing of the Binder ratio for various system sizes via Monte-Carlo (red squares). $p_c^{(n)}$ in the replica limit $n \to 1$ (the yellow star) is given by the critical point of random-bond Ising model (RBIM) in 2D, $p_c^{(1)} \approx 0.109$, as explained in Sec.~\ref{sec:duality}. In the limit $n\to \infty$, the spin model is asymptotically decoupled Ising models with $p_c^{(\infty)}\approx 0.293$ (the grey dashed line).}
    \label{fig:critical}
\end{figure}

Here, we study the ferromagnetic transition in the $(n-1)$-flavor Ising model.
The transition points depend on $n$ and are determined using both analytical methods (e.g. Kramers-Wannier duality for $n = 2,3$) and Monte-Carlo simulation (for $n = 4,5,6$, etc).
The results are presented in Fig.~\ref{fig:critical}.

For $n = 2$, the statistical mechanical model is the standard square lattice Ising model:
\begin{equation}
    H_{2,a} = -2J_a\sum_{\langle i,j\rangle} \sigma_i \sigma_j\,.
\end{equation}
The critical point is determined analytically by the Kramers-Wannier duality~\cite{potts1952some,kihara1954statistics}
\begin{align}
    p_{c}^{(2)} = \frac{1}{2}\Big(1 - \sqrt{\sqrt{2}-1}\,\Big) \approx 0.178.
\end{align}

For $n = 3$, the model becomes the Ashkin-Teller model on $2$D square lattice along the $\mathcal{S}_4$ symmetric line. 
The Hamiltonian is
\begin{equation}
\label{eq:n=3 Hamiltonian}
    H_{3,a} = -J_a\sum_{\langle i,j\rangle}\sigma_{i}^{(1)}\sigma_j^{(1)} + \sigma_i^{(2)}\sigma_j^{(2)}
    + \sigma_i^{(1)}\sigma_i^{(2)}\sigma_j^{(1)}\sigma_j^{(2)}.
\end{equation}
The model is equivalent to the standard four-state Potts model~\cite{kohmoto1981hamiltonian} with a critical point determined by the Kramers-Wannier duality
\begin{equation}
    p_{c}^{(3)} = \frac{1}{2}\Big(1 - \frac{1}{\sqrt{3}}\Big) \approx 0.211.
\end{equation}

\begin{figure}
    \centering
    \includegraphics[width=0.46\textwidth]{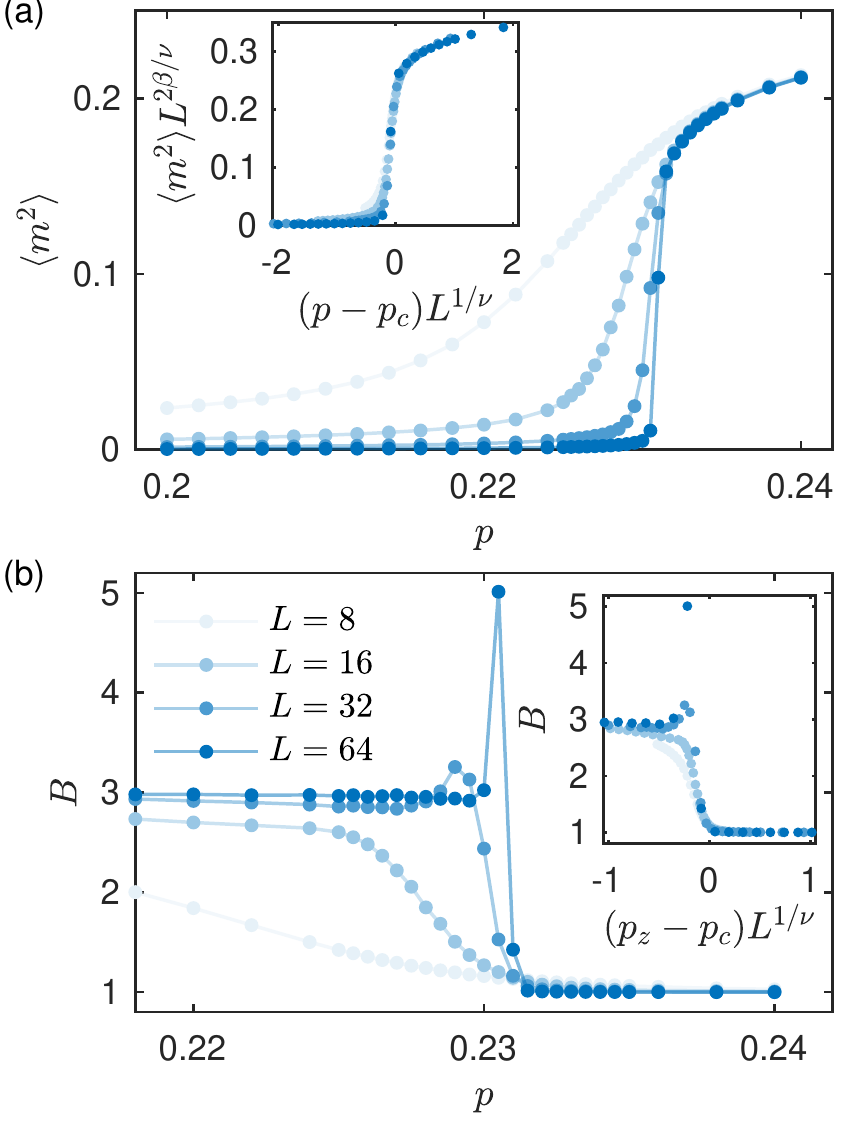}
    \caption{Phase transition in the statistical mechanical model for $n = 4$. Magnetization (a) and Binder ratio (b) as a function of error rate $p$ for various system sizes up to $L_x = L_y = L = 64$. The crossing of $B(p,L)$ yields $p_c = 0.231$. The exponents $\nu = 0.74$ and $\beta = 0.04$ are extracted from the finite-size scaling collapse in the insets. The results are averaged over $10^5$ independent Monte-Carlo measurements for each of $48$ initial spin configurations.}
    \label{fig:transition}
\end{figure}

For $n \geq 4$, we are not aware of any exact solution and resort to the Monte-Carlo simulation.
To locate the transition point $p_c$, we consider the average magnetization per spin,
\begin{equation}
	m := \frac{1}{(n-1)L^2}\sum_{s = 1}^{n-1} \sum_{i} \sigma^{(s)}_i.
\end{equation}
We calculate the magnetization square $\langle m^2 \rangle$ and the Binder ratio $B = \langle m^4 \rangle/\langle m^2 \rangle^2$ numerically and display the results in Fig.~\ref{fig:transition}.
Assuming a continuous transition, we determine $p_c^{(n)}$ by the crossing point of $B(p, L)$ for various system sizes $L$ and extract the critical exponents using the scaling ansatz $B(p, L) = \mathcal{F}_b((p-p_c)L^{1/\nu})$ and $\langle m^2 \rangle (p, L) = L^{-2\beta/\nu}\calF_m((p-p_c)L^{1/\nu})$.
The analysis yields $p_c^{(4)} = 0.231$ for $n = 4$.
However, the sharp drop of magnetization and the non-monotonic behavior of $B(p,L)$ near $p_c^{(4)}$ hint at a possible first-order transition~\cite{Binder:1984llk,Iino:2018}.

The critical error threshold $p_c$ increases monotonically with $n$ and is exactly solvable in the limit $n \to \infty$.
In this case, the interaction among different flavors is negligible compared to the two-body Ising couplings.
Thus, the critical point is asymptotically the same as that in the Ising model with coupling $J_a$ and is given by
\begin{equation}
    p_{a,c}^{(\infty)} = \frac{1}{2}\big(2-\sqrt{2}\, \big) \approx 0.293.
\end{equation}

\subsection{Three diagnostics}\label{sec:stat-mech diagnostics}

 The R\'enyi version of the three information-theoretic diagnostics, quantum relative entropy, coherent information, and topological entanglement negativity, translate into distinct physical quantities in the statistical mechanical model. We write these quantities explicitly below and show that all three detect the establishment of ferromagnetic order. Therefore the transition in all three quantities is governed by the same critical point, a fact that is not evident before mapping to statistical mechanical models.

\subsubsection{Quantum relative entropy}

We start with the R\'enyi version of the quantum relative entropy given by \eqnref{eq:renyi_rel}. Let $\rho$ be the corrupted ground state of the Toric code, and $\rho_m = \calN[\ket{\Psi_m} \bra{\Psi_m}]$ where $\ket{\Psi_m} := w_m(\mathcal{C})\ket{\Psi_0}$ has a pair of $m$-particles at the end of path $\mathcal{C}$.
The phase errors do not change the distinguishability between the two states and can be safely ignored here.
Only the statistical mechanical model for the $Z$ loops/spins is relevant.
Let $i_\ell$ and $i_r$ denote the positions of two $m$-particles, we show in \appref{sec:stat-mech_rel} that the R\'enyi relative entropy is mapped to a two-point function of the Ising spins
\begin{align}
	D^{(n)}(\rho||\rho_\alpha) = \frac{1}{1-n}\log\langle \sigma_{i_\ell}^{(1)} \sigma_{i_r}^{(1)} \rangle\,,
\end{align}
where $\sigma^{(1)}_j$ is the first flavor of the Ising spin at site $j$, and the subscription $z$ is suppressed.

When the error rate is small and the system is in the paramagnetic phase, the correlation function decays exponentially, and thus $D^{(n)} = O(|i_\ell - i_r|)$ which grows linearly with the distance between $i_\ell$ and $i_r$. 
This indicates that the error-corrupted ground state and excited state remain distinguishable.
When the error rate exceeds the critical value and the system enters the ferromagnetic phase, $D^{(n)}$ is of $O(1)$ due to the long-range order, which implies that the error-corrupted ground state and excited state are no longer distinguishable.

\subsubsection{Coherent information}

Next consider the R\'enyi version of the coherent information $I_c^{(n)}$ in \eqnref{eq:renyi_Ic}.
We let the two logical qubits in the system $Q$ be maximally entangled with two reference qubits $R$.
As detailed in \appref{sec:stat-mech_Ic}, $I_c^{(n)}$ can be mapped to the free energy cost of inserting domain walls along non-contractible loops that are related to the logical operators.
More explicitly, let $\mathbf{d}_{al}$ with $a=x,z$ and $l = l_1,l_2$ be a $(n-1)$-component binary vector.
Each component of $\mathbf{d}_{al}$ dictates the insertion of domain walls for $a = x,z$ spins along the non-contractible loop $l$, respectively, in $n-1$ copies of the Ising spins.
Here, along the domain walls, the couplings between nearest neighbor spins are flipped in sign and turned anti-ferromagnetic.
Then, we have
\begin{equation}
\label{eq:coherent information mapping}
	I_c^{(n)} = \frac{1}{n-1} \sum_{a = x,z} \log \Big( \sum_{\mathbf{d}_{a1}\mathbf{d}_{a2}} e^{-\Delta F_{n,a}^{(\mathbf{d}_{a1},\mathbf{d}_{a2})}} \Big) - 2\log 2\,,
\end{equation}
where $\Delta F_{n,a}^{(\mathbf{d}_{a1},\mathbf{d}_{a2})}$ is the free energy cost associated with inserting domain walls labeled by binary vectors $\mathbf{d}_{al}$, the sum runs over all possible $\mathbf{d}_{al}$.

When the error rate is small and the system is in the paramagnetic phase, the domain wall along a non-contractible loop costs nothing, i.e. $\Delta F_{n,a}^{(\mathbf{d}_{a1},\mathbf{d}_{a2})} = 0$.
It follows that the corrupted state retains the encoded information, i.e. $I_c^{(n)} = 2\log 2$.
When the error rate exceeds the critical value and the system enters the ferromagnetic phase, inserting a domain wall will have a free energy cost that is proportional to its length. 
Namely, $\Delta F_{n,a}^{(\mathbf{d}_{a1},\mathbf{d}_{a2})}$ is proportional to the linear system size unless no defect is inserted.
One can deduce $I_c^{(n)} = 0$ when the spin model for either $Z$ or $X$ loop undergoes a transition to the ferromagnetic phase, namely, the corrupted state corresponds to a classical memory.
When both spin models are in the ferromagnetic phase, we have $I_c^{(n)} = -2\log 2$, indicating that the system is a trivial memory.

\subsubsection{Topological entanglement negativity}

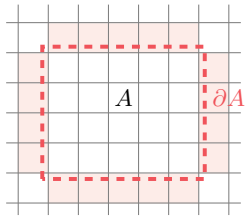
\begin{figure}[t!]
    \centering
    \begin{tikzpicture}[scale=0.8]
    \definecolor{myred}{RGB}{240,83,90};
    \definecolor{myblue}{RGB}{93,123,189};
    \definecolor{mylightred}{RGB}{252,218,210};
    \definecolor{myturquoise}{RGB}{83,195,189};
    \foreach \y in {0,...,3}
    {\filldraw[mylightred, opacity=0.5] (-0.5, 0.5*\y+1) rectangle (-0.5+0.48,0.5*\y+1+0.48);}
    \foreach \y in {0,...,3}
    {\filldraw[mylightred, opacity=0.5] (2.5, 0.5*\y+1) rectangle (2.5+0.48,0.5*\y+1+0.48);}
    \foreach \x in {0,...,4}
    {\filldraw[mylightred, opacity=0.5] (0.5*\x, 0.5) rectangle (0.5*\x+0.48,0.5+0.48);}
    \foreach \x in {0,...,4}
    {\filldraw[mylightred, opacity=0.5] (0.5*\x, 3) rectangle (0.5*\x+0.48,3+0.48);}
    \foreach \x in {-1,0,...,6}
    {\draw[gray] (0.5*\x, 0.3) -- (0.5*\x, 3.8);}
    \foreach \y in {0,...,6}
    {\draw[gray] (-0.7,0.5+0.5*\y) -- (3.2,0.5+0.5*\y);}
    \draw[dashed,myred,line width=1.5] (-0.1,0.9) -- (-0.1,3.1) -- (2.6,3.1) -- (2.6,0.9) -- (-0.1,0.9);
    \node[myred] at (3.0,2.25) {$\partial A$};
    \node at (1.25,2.25) {$A$};
    \end{tikzpicture}
    \caption{Entanglement negativity between region $A$ and its complement $\bar{A}$ corresponds to the excess free energy for aligning Ising spins on the boundary of $A$ (pink plaquettes) pointing to the same direction.}
    \label{fig:neg_constraint}
\end{figure}

The R\'enyi negativities of even order are given in \eqnref{eq:renyi_neg}. Let us specialize here to the Toric code with only phase errors.
As shown in \appref{sec:stat-mech_neg}, the $2n$-th R\'enyi negativity of a region $A$ is given by
\begin{align}
	\mathcal{E}^{(2n)}_{A} = \Delta F_A\,,
\end{align}
where $\Delta F_A$ is the excess free energy associated with aligning a single flavor of Ising spins on the boundary $\partial A$ in the same direction (illustrated in \figref{fig:neg_constraint}).  

The excess free energy $\Delta F_A$, or more precisely, its subleading term can probe the ferromagnetic transition in the statistical-mechanical model.
The excess free energy has two contributions.
The energetic part is always proportional to $|\partial A|$.
The entropic part is attributed to the loss of degrees of freedom due to the constraint.
In the paramagnetic phase, the Ising spins fluctuate freely above the scale of the finite correlation length $\xi$.
Hence, enforcing each constraint removes $O(|\partial A|/\xi)$ degrees of freedom proportional to the circumference of $A$, which yields the leading term (area law).
Importantly, there is still one residual degree of freedom, namely, the aligned boundary spins can fluctuate together, which results in a subleading term $\log 2$.
Altogether, we have $\mathcal{E}^{(2n)}_{A} = c|\partial A|/\xi - \log 2$. 
Here, it is an interesting question to verify whether the prefactor $c$ is universal or not~\cite{Metlitski:2009iyg}, and we leave it for future study~\footnote{We thank Tarun Grover for pointing it out to us.}.
In the ferromagnetic phase, the finite correlation length $\xi$ sets the scale of the critical region, below which the spins can fluctuate.
Thus, imposing each constraint removes $O(|\partial A|/\xi)$ degrees of freedom.
However, the aligned boundary spins should also align with the global magnetization resulting in a vanishing subleading term in the excess free energy.
Hence, the negativity $\mathcal{E}_{A}^{(2n)}$ exhibits a pure area law without any subleading term.

\begin{figure}[t]
    \centering
    \includegraphics[width=0.46\textwidth]{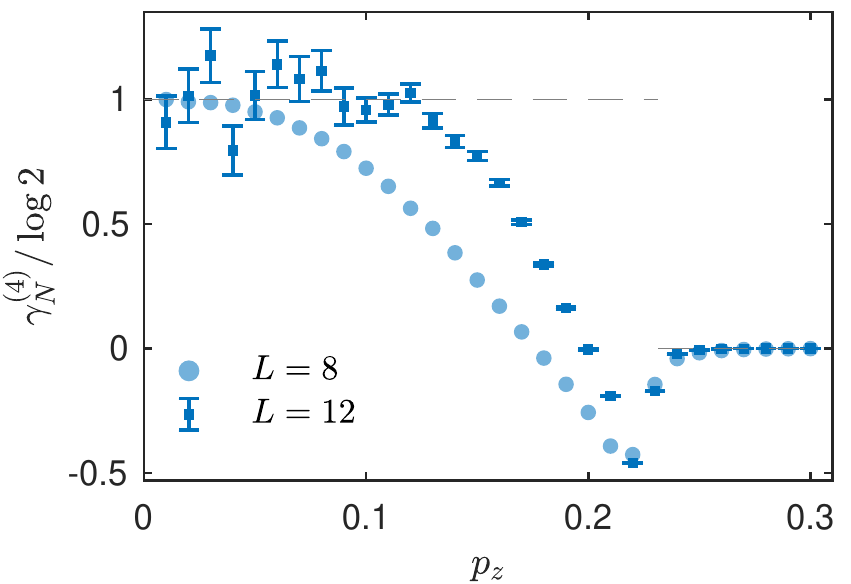}
    \caption{Topological negativity $\gamma_N^{(4)}$ as a function of the phase error rate $p_z$. 
    We consider the subsystems $A, B,$ and $C$ as in Eq.~\eqref{eq:gammaN} and choose the side of the region $ABC$ to be $L/4$.
    $\gamma_N^{(4)}$ approaches $\log 2$ and zero at small and large $p_z$, respectively. 
    The curves become steeper as the system size $L$ increases. 
    The dashed line indicates the predicted behavior in the thermodynamic limit. 
    The results are averaged over $10^7$ independent Monte-Carlo measurements from each of $48,96$ random initial spin configurations for $L = 8, 12$, respectively.
    The error bars for $L = 8$ are negligible and thus omitted.
    }
    \label{fig:gamma}
\end{figure}

To support our analytical argument, we also numerically calculate the R\'enyi-$4$ negativity (the R\'enyi-$2$ negativity is trivially zero) and show that the topological term $\gamma_N^{(4)}$ indeed exhibits distinct behaviors across the transition.
We adopt the Kitaev-Preskill prescription to extract $\gamma_N$~\cite{Kitaev:2005dm}. 
More specifically, we consider the subsystems $A$, $B$, $C$ depicted below, and $\gamma_N$ is given by
\begin{equation}
\begin{tikzpicture}[scale=0.6, baseline={(current bounding box.center)}]
\definecolor{myred}{RGB}{240,83,90};
\definecolor{myblue}{RGB}{93,123,189};
\definecolor{myturquoise}{RGB}{83,195,189};
\small
\foreach \y in {0.7,1.4,...,4.2}{
	\draw[black!40] (0.7-0.2,\y) -- (4.4,\y);
}
\foreach \x in {0.7,1.4,...,4.2}{
	\draw[black!40] (\x,0.7-0.2) -- (\x,4.4);
}
\draw[myturquoise, line width=2] (1.4,1.4) -- (3.5,1.4) -- (3.5,3.5) -- (3.5,2.8) -- (2.8,2.8) -- (2.8,1.4) -- (2.1,1.4) -- (2.1,2.1) -- (3.5,2.1);
\draw[myred, line width=2] (1.4,3.5) -- (3.5,3.5) -- (2.8,3.5) -- (2.8,2.8) -- (2.1,2.8) -- (2.1,3.5);
\draw[myblue, line width=2] (1.4,1.4) -- (1.4,3.5) -- (1.4,2.8) -- (2.1,2.8) -- (2.1,2.1) -- (1.4,2.1);
\node[myred] at (2.45,3.85) {$A$};
\node[myblue] at (1.05,2.45) {$B$};
\node[myturquoise] at (3.85,1.05) {$C$};
\end{tikzpicture}\; \; 
\begin{aligned}
-\gamma_N := \;&\calE_A + \calE_B + \calE_C + \calE_{ABC}\\
&-\calE_{AB} -\calE_{BC} -\calE_{AC}\,.
\end{aligned}\label{eq:gammaN}
\end{equation}
Our choice of the subsystems further simplifies the above expression to $-\gamma_N = 2\calE_A - 2\calE_{AC} + \calE_{ABC}$~\footnote{Obtaining the negativity from the Monte-Carlo simulation is not an easy task. 
Here, one directly computes $e^{(2-n)\calE_A^{(n)}}$, which is exponentially small due to the area-law scaling of $\calE_A^{(n)}$ and thus requires exponentially many samples to accurately determine its value.
This limits the largest accessible subsystem size.}.

The result is presented in Fig.~\ref{fig:gamma}, where $\gamma_N^{(4)}$ approaches $\log 2$ and $0$ for small and large $p_z$, respectively.
The curves become steeper as the system size increases, which is consistent with the predicted step function in the thermodynamic limit.
One can also observe a dip of $\gamma_N^{(4)}$ below zero.
This phenomenon has also appeared in the numerical study of the topological entanglement entropy across transitions~\cite{Verresen:2020dmk}.
We believe that this dip is due to the finite-size effect, which might be more severe for information quantities with a large R\'enyi index $n$~\cite{Jiang:2013pwa}.

So far, we only considered a simply connected sub-region. 
If $A$ is not simply connected, that is, $\partial A$ contains $k$ disconnected curves (for example the boundary of an annular region that contains two disconnected curves), then the constraints only require the Ising spins to align with other spins on the same boundary curve. 
In this case, the topological entanglement negativity is $k\log 2$.
This is the same dependence on the number of disconnected components as in the topological entanglement entropy of ground states~\cite{LevinWen2005}.

\subsection{$n\rightarrow 1$ limit, duality and connection to optimal decoding}
\label{sec:duality}

In this subsection, we determine $p_c$ in the limit $n\rightarrow 1$ via a duality between the statistical mechanical model established in \secref{sec:stat-mech model} and the 2D random bond Ising model (RBIM) along the Nishimori line~\cite{nishimori1981internal}.
The RBIM is also known to govern the error threshold of the optimal decoding algorithm for the 2D Toric code with incoherent errors~\cite{Dennis:2001nw}.
The duality shows that the decoding threshold indeed saturates the upper bound given by the threshold in our information theoretical diagnostics.
This duality was derived before via a binary Fourier transformation~\cite{NishimoriDuality,OhzekiDuality}. Here, it follows naturally from two distinct expansions of the error-corrupted state.

The statistical mechanical model in \secref{sec:stat-mech model} is based on the loop picture in Eq.~\eqref{eq:rho0_loop}.
Here, we work in an alternative error configuration picture, writing the error corrupted state as
\begin{equation}
	\begin{tikzpicture}[scale=0.75, baseline={(current bounding box.center)}]
		\definecolor{myred}{RGB}{240,83,90};
		\definecolor{myblue}{RGB}{73,103,189};
		\small
		\foreach \y in {0,0.5,...,1.5,2}{
			\draw[black!60] (-0.2,\y) -- (2.2,\y);
		}
		\foreach \x in {0,0.5,...,2}{
			\draw[black!60] (\x,-0.2) -- (\x,2.2);
		}
		\draw[myred,line width=1] (0,0) -- (0,1) -- (1,1);
		\draw[myred,line width=1] (0.5,2) -- (1,2) -- (1,1.5) -- (0.5,1.5) -- (0.5,2);
		\draw[myblue,line width=1.5, dashed] (1.75,0.25) -- (1.75,1.25);
		\filldraw[myred] (1,1) circle (0.1) node[below left, red]{\scriptsize$e$};
		\filldraw[myred] (0,0) circle (0.1) node[below right, red]{\scriptsize$e$};
		\filldraw[myblue] (1.75,0.23) circle (0.1) node[below, myblue]{\scriptsize$m$};
		\filldraw[myblue] (1.75,1.27) circle (0.1) node[above, myblue]{\scriptsize$m$};
		\node[myred] at (-0.27,0.75) {$\calC_z$};
		\node[myblue] at (2.25,0.75) {$\calC_x$};
	\end{tikzpicture} \qquad
\begin{aligned}
	\rho = & \sum_{\mathcal{C}_x, \mathcal{C}_z} P(\mathcal{C}_x) P(\mathcal{C}_z) \\
	& \qquad Z^{ \mathcal{C}_z} X^{ \mathcal{C}_x} \rho_0 X^{ \mathcal{C}_x}Z^{ \mathcal{C}_z}\,,
\end{aligned}
\end{equation}
where $\mathcal{C}_z$ ($\mathcal{C}_x$) denotes the error string on the original (dual) lattice. The error string creates error syndromes, i.e. $e$ ($m$) anyons, on the boundary $\partial\mathcal{C}_{z}$ ($\partial\mathcal{C}_{x}$). 
The probability for each string configuration is
\begin{equation}
	P(\mathcal{C}_{a}) = p_a^{|\calC_a|}(1-p_a)^{N-|\calC_a|}\,,
\end{equation}
where $|\calC_a|$ with $a = x,z$ denotes the total length of the error string, and $N$ is the total number of qubits.

The expansion in error configurations allows writing the $n$-th moment as
\begin{equation}
	\begin{aligned}
		\tr \rho^n = &\sum_{\{\mathcal{C}^{(s)}_{x},\;\mathcal{C}^{(s)}_{z}\}} \prod_{s=1}^n P\big(\mathcal{C}^{(s)}_{x}\big)
		P\big(\mathcal{C}^{(s)}_{z}\big) \\
		&\quad \tr \Big(\prod_{s = 1}^n Z^{ \mathcal{C}^{(s)}_z}X^{ \mathcal{C}^{(s)}_{x}} \rho_0 X^{ \mathcal{C}^{(s)}_x}Z^{ \mathcal{C}^{(s)}_z}\Big).
	\end{aligned}\label{eq:trrho_error}
\end{equation}
We again choose $\rho_0$ to be the maximally mixed state in the logical space.
The trace is non-vanishing only when the error strings in two consecutive copies differ by a closed loop.
Thus, the error strings in the $2,\ldots,n$-th copies are related to that in the first copy via 
\begin{equation}
\begin{tikzpicture}[scale=0.75, baseline={(current bounding box.center)}]
	\definecolor{myred}{RGB}{240,83,90};
	\definecolor{myblue}{RGB}{73,103,189};
	\small
	\foreach \y in {0,0.5,...,1.5,2}{
		\draw[black!60] (-0.2,\y) -- (2.2,\y);
	}
	\foreach \x in {0,0.5,...,2}{
		\draw[black!60] (\x,-0.2) -- (\x,2.2);
	}
	\draw[myred,line width=1.05] (0,0.5) -- (0,1.5) -- (1,1.5);
	\draw [dashed, line width=1.05, draw=myred, fill=gray, fill opacity=0.2]
	(0,0.5) -- (0,1.5) -- (1,1.5) -- (1,2) -- (1.5,2) -- (1.5,1) -- (1,1) -- (1,0.5) -- cycle;
	\draw [dashed, line width=1.05, draw=myred, fill=gray, fill opacity=0.2]
	(1.5,0) -- (1.5,0.5) -- (2,0.5) -- (2,0) -- cycle;
	\draw[->,>=stealth] (2.3,1.1) -- (1.25,1.25);
	\draw[->,>=stealth] (2.3,0.8) -- (1.75,0.25);
	\node[black] at (2.75,1.05) {$v_z^{(s)}$};
	\filldraw[myred] (1,1.5) circle (0.1) node[above left, red]{\scriptsize$e$};
	\filldraw[myred] (0,0.5) circle (0.1) node[below right, red]{\scriptsize$e$};
	\node[myred] at (-0.36,1.1) {$\calC_z^{(1)}$};
\end{tikzpicture} \,\,
\begin{gathered}
	\mathcal{C}^{(s+1)}_{a} = \mathcal{C}^{(1)}_{a} + \partial v^{(s)}_{a} + l_1^{d^{(s)}_{a1}} + l_2^{d^{(s)}_{a2}}, \\ 
	s = 1,\ldots, n-1\,,
\end{gathered}
\end{equation}
where $v_{a}^{(s)}$ is a set of plaquettes on the original (or dual) lattice, its boundary $\partial v^{(s)}_{a}$ only consists of homologically trivial loops. Two strings can also differ by a non-contractible loop $l_1, l_2$ on the torus indicated by the binary variables $d^{(s)}_{a1},d^{(s)}_{a2} = 0,1$.
Noticing the decoupling between $Z$ and $X$, we have
\begin{equation}
\begin{aligned}
    \tr \rho^n =& \, \calZ_{n,z}' \calZ_{n,x}'\,, \\
    \calZ_{n,a}' =& \sum_{\mathbf{d}_{a}}\sum_{\calC_{a}^{(1)}} P\big( \calC_{a}^{(1)} \big) \\
    &\sum_{\{ v_{a}^{(s)} \}} \prod_{s=1}^{n-1} P\big( \mathcal{C}^{(1)}_{a} + \partial v^{(s)}_{a} + l_1^{d^{(s)}_{a1}} + l_2^{d^{(s)}_{a2}}\big)\,,
\end{aligned}
\end{equation}
where we denote the collection of $(d_{a1}^{(s)}, d_{a2}^{(s)})$ for $s = 1,2,\cdots,n-1$ as a binary vector $\mathbf{d}_{a}$.
Comparing the above expression with Eq.~\eqnref{eq:factorization}, we have
\begin{equation}
    \calZ_{n,x} = 2^{\frac{(n-1)N}{2}} \calZ_{n,z}'\,,\quad
    \calZ_{n,z} = 2^{\frac{(n-1)N}{2}} \calZ_{n,x}'\,.
\end{equation}
In the following, we focus on $\calZ_{n,z}'$ and suppress the subscripts for clarity. The analysis of $\calZ_{n,x}'$ is similar.

We now interpret $\calZ_{n}'$ as a partition function of Ising spins that is related to the replicated RBIM.
We first introduce $n-1$ flavors of Ising spins on the plaquettes to represent $v^{(s)}, s=1,\ldots,n-1$. The Ising domain wall represents $\partial v^{(s)}$ as shown below.
\begin{equation*}
\begin{tikzpicture}[scale=0.8, baseline={(current bounding box.center)}]
	\definecolor{myred}{RGB}{240,83,90};
	\definecolor{myblue}{RGB}{73,103,189};
	\small
	\foreach \y in {0,0.5,...,1.5,2}{
		\draw[black!60] (-0.2,\y) -- (2.2,\y);
	}
	\foreach \x in {0,0.5,...,2}{
		\draw[black!60] (\x,-0.2) -- (\x,2.2);
	}
	\foreach \x in {0,0.5,1}{
		\draw[->,>=stealth] (\x+0.25,0.1) -- (\x+0.25,0.4);
	}
	\foreach \x in {0,0.5,1.5}{
		\draw[->,>=stealth] (\x+0.25,0.1+1.5) -- (\x+0.25,0.4+1.5);
	}
	\draw[->,>=stealth] (1.25,0.1+0.5) -- (1.25,0.4+0.5);
	\draw[->,>=stealth] (1.75,0.1+0.5) -- (1.75,0.4+0.5);
	\draw[->,>=stealth] (1.75,0.1+1) -- (1.75,0.4+1);
	\foreach \x in {0,0.5,1}{
		\draw[<-,>=stealth] (\x+0.25,0.1+1) -- (\x+0.25,0.4+1);
	}
	\foreach \x in {0,0.5}{
		\draw[<-,>=stealth] (\x+0.25,0.1+0.5) -- (\x+0.25,0.4+0.5);
	}
	\draw[<-,>=stealth] (1.75,0.1) -- (1.75,0.4);
	\draw[<-,>=stealth] (1.25,0.1+1.5) -- (1.25,0.4+1.5);
	\draw[myred,line width=1.05] (0,0.5) -- (0,1.5) -- (1,1.5);
	\draw [dashed, line width=1.05, draw=none, fill=gray, fill opacity=0.2]
	(0,0.5) -- (0,1.5) -- (1,1.5) -- (1,2) -- (1.5,2) -- (1.5,1) -- (1,1) -- (1,0.5) -- cycle;
	\draw [dashed, line width=1.05, draw=none, fill=gray, fill opacity=0.2]
	(1.5,0) -- (1.5,0.5) -- (2,0.5) -- (2,0) -- cycle;
	\filldraw[myred] (1,1.5) circle (0.1);
	\filldraw[myred] (0,0.5) circle (0.1);
\end{tikzpicture} 
\end{equation*}
Next, we identify the probability of error strings with the Boltzmann weight of Ising spin configurations.
Effectively, the spins of the same flavor have nearest neighbor anti-ferromagnetic interactions if their link crosses the path $\calC^{(1)}$ or $l_{1(2)}$ when ${d_{1(2)}^{(s)}} = 1$; the interaction is ferromagnetic otherwise.
Specifically,
\begin{equation}
\begin{aligned}
    \calZ_{n}' =& ((1-p)p)^{N/2} \sum_{\{ \eta_{ij} \}} P(\{ \eta_{ij} \}) \sum_{\mathbf{d}}\sum_{\tau^{(s)}} e^{-\mathsf{H}_n(\eta_{ij}, \mathbf{d})},
\end{aligned}
\end{equation}
where
\begin{equation}
    \mathsf{H}_n(\eta_{ij}, \mathbf{d}) =  -J \sum_{s=1}^{n-1} \sum_{\braket{ij}} \xi_{ij}^{(s)}(\mathbf{d}) \eta_{ij} \tau_i^{(s)} \tau_j^{(s)}.
\end{equation}
Here, $J$ depends on $p$ and satisfies the Nishimori condition $e^{-2J} = p/(1-p)$~\cite{nishimori1981internal,Dennis:2001nw}. 
Both $\eta_{ij}$ and $\xi_{ij}^{(s)}(\mathbf{d})$ take the value $\pm 1$.
The random variable $\eta_{ij}$ takes the $-1$ value along $\calC^{(1)}$, which can be interpreted as a random sign of bond coupling. 
The variable $\xi_{ij}^{(s)}(\mathbf{d})=-1$ along non-contractible loops $l_{1(2)}$ when $d_{1(2)}^{(s)} = 1$, which can be interpreted as a defect in the spin model. 
%
The above expression allows writing $\mathcal{Z}'_{n} = \overline{\mathcal{Z}_{\text{RBIM}}^{n-1}}$ as the disorder-averaged partition function of $n-1$ copies of RBIM along the Nishimori line.

The replicated RBIM in the error configuration picture and the spin model in the loop picture are derived from the $n$-th moment of the \emph{same} error corrupted state. 
Therefore, they are dual to each other and share the same critical error rate for all replica indices.
Note that the replicated RBIM exhibits two phases, a ferromagnetic and a paramagnetic phase at small and large error rates, respectively.
The phase diagram is exactly opposite to that of the spin model in its dual picture, which is a common feature of Kramers-Wannier dualities~\footnote{One can show the usual Kramers-Wannier duality between the two spin models by performing high- and low- temperature expansions.}.

In the replica limit $n \to 1$, the replicated RBIM reduces to the RBIM derived for the optimal quantum error correction algorithm~\cite{Dennis:2001nw} and undergoes an ordering transition at $p_c = 0.109$~\cite{honecker2001universality}\footnote{We note that the replicated RBIM in the limit $n \to 1$ contains a summation over different boundary conditions. However, the critical point does not depend on the boundary condition as the ferromagnetic transition is a bulk transition.}.
This implies that all three diagnostics should also undergo the transition at the same $p_c$ in the replica limit and confirms that the optimal decoding threshold saturates the upper bound in Eq.~\eqref{eq:pc_upper}.

\section{Discussion}
\label{sec:discussion}

In this work, we introduced information theoretic diagnostics of  error-corrupted mixed states $\rho = \prod_{i} \calN_i[ \rho_0]$, which probe their intrinsic topological order and capacity for protecting quantum information. We focused on a concrete example, where $\rho_0$ is in the ground state subspace of the Toric code and $\calN_i$ describes the bit-flip and phase errors.
We noted that the $n$-th moment $\tr \rho^n$ can be written as the partition function of a 2D classical spin model, that is dual to the (replicated) random-bond Ising model along the Nishimori line, which is used to establish the following results.
We consider three complementary diagnostics, quantum relative entropy, coherent information, and topological entanglement negativity, which are mapped to different observables in the spin model and shown to undergo a transition at the same critical error rate. 
Generally speaking, this critical error rate is an upper bound for the error threshold that can be achieved by any decoding algorithm.
The aforementioned duality implies that the critical error rate identified here is exactly saturated by the error threshold of the optimal decoding algorithm for the Toric code proposed by Dennis et al~\cite{Dennis:2001nw}.
This result unveils a connection between the breakdown of topological quantum memory and a transition in the mixed-state topological order, and also provides physical interpretation for the decoding transition.

We have focused on Toric code with incoherent errors.
It will be interesting to generalize the discussion to coherent errors that create anyons with coherence, e.g., amplitude damping or unitary rotations~\cite{Poulin2017,Poulin2018,Bravyi2018Coherent,Venn:2022kxy}.
In these cases, one has to concatenate coherent errors and dephasing channels that mimic the syndrome measurement in order to make better contact to quantum error correction based on that syndrome measurement.
It is also interesting to further consider non-Abelian quantum codes~\cite{Brell:2013wsa,Wootton2014,Schotte:2020lnz}.

It might be surprising that the intrinsic properties of the 2D error corrupted quantum states are captured by 2D {\em classical} statistical mechanical models. 
In \appref{app:Zn toric code}, we give a brief discussion on $\bbZ_N$ Toric code with specific incoherent errors and show that this is also the case.
A more general perspective is the so-called \emph{errorfield double formalism}, which is proposed by the same authors.
It follows from this general formalism that the intrinsic properties of the 2D error corrupted states can always be captured by a 1+1D quantum model.
Details will be reported elsewhere~\cite{EFDtoappear}.

In the Toric code and other topological codes with local errors, the statistical mechanical model for the optimal decoding algorithm always satisfies the Nishimori condition~\cite{Dennis:2001nw,Wang:2002ph,Katzgraber:2009zz,Bombin:2012jk,Kubica:2018rab,Flammia2021,Song:2021bud}.
One salient feature of the statistical mechanical model on the Nishimori line is an enlarged $\calS_{n}$ symmetry in the replicated model of $n-1$ replicas~\cite{le1988location,Ludwig2001PRB}.
%
In our analysis of intrinsic mixed-state transition, the $(n-1)$-th replicated model actually corresponds to the $n$-th moment $\tr\rho^n$, where the invariance under permuting $n$ copies of the density matrix naturally gives rise to the $\calS_n$ symmetry. This offers an alternative perspective on the occurrence of Nishimori physics in the context of optimal decoding.


As we have commented in \secref{sec:error corrupted state}, the error-induced transition acquires a different nature from the thermal transition in finite-temperature topological order.
This distinction suggests a hierarchy of topological transitions in general mixed states.
For example, it suffices to use physical observables (linear in the density matrix) to detect the thermal transition, while it requires at least second R\'enyi quantities (quadratic in the density matrix) to detect the error-induced transition.
It is interesting to explore more exotic topological transitions in mixed states that are detectable only by non-linear functions of the density matrix of even higher orders, such as the entanglement Hamiltonian.

The above task is intimately related to the goal of classifying mixed-state topological order.
A suitable definition of mixed state topological order should be both operationally meaningful and also identify computable topological invariants. 
Our discussion which focuses on the error-corrupted mixed states represents one particular aspect of this more general question.
Here, the coherent information provides the operational definition, namely, a locally corrupted state is in a different phase if QEC is impossible, while the topological entanglement negativity is believed to provide a computable topological invariant that diagnoses the present transition. 
However, note that both the local error channel and QEC process are generally non-unitary, for which the Lieb-Robinson bound does not apply. 
Therefore, understanding the role of locality is key to obtaining a general notion of equivalence classes of mixed states. Similarly, a more general justification of topological negativity and its universality, in the sense of establishing its  invariance under the application of local quantum channels at a certain place, is left for future work. 
The main difficulty comes from understanding how local perturbations affect the spectrum of a partially transposed density matrix, which is an interesting problem in its own right and is left to future work.

\begin{acknowledgments}
We thank Meng Cheng, Soonwon Choi, Mikhail Lukin, Nishad Maskara, Karthik Siva, Tomohiro Soejima for helpful discussions, and Tarun Grover for useful comments on the manuscript. AV was funded by the Simons Collaboration on Ultra-Quantum Matter, which is a grant from the Simons Foundation (651440, AV). AV and RF further acknowledge support from NSF-DMR 2220703. Support is also acknowledged from the U.S. Department of Energy, Office of Science, National Quantum Information Science Research Centers, Quantum Systems Accelerator (EA). YB was supported in part by NSF QLCI program through grant number OMA-2016245. This work is funded in part by a QuantEmX grant from ICAM and the Gordon and Betty Moore Foundation through Grant GBMF9616 to Ruihua Fan and Yimu Bao. This research used the Savio computational cluster resource provided by the Berkeley Research Computing program at the University of California, Berkeley.
\end{acknowledgments}

\emph{Note added}: Upon completion of the present manuscript, we became aware of an independent work~\cite{CenkeToAppear} which is broadly related and will appear on arXiv on the same day. We thank them for informing us their work in advance.

\appendix
\section{Details of the mapping}
\label{sec:details}

In this section, we detail the mapping between the three diagnostics and observables in the statistical mechanical models.

\subsection{Quantum relative entropy}
\label{sec:stat-mech_rel}
We here explicitly show that the R\'enyi quantum relative entropy is related to the correlation function in the classical spin model.
Specifically, we consider the relative entropy between the error corrupted ground state and an excited state $\ket{\Psi_m} := w_m(\mathcal{C})\ket{\Psi_0}$ with a pair of $m$-particles created at the end of path $\mathcal{C}$.

First, we write down the error corrupted state $\rho_m$ in the loop representation
\begin{align}
    \rho_m &= \frac{1}{2^N} \sum_g \sgn\big(g_z, X^{\mathcal{C}}\big) g_z g_x e^{-\mu_x|g_x|-\mu_z|g_z|}.
\end{align}
where the commutation relation between the loop operator and the string operator is accounted by $\sgn(g_z, X^{\mathcal{C}})$; the sign function equals $+1$ when $g_z$ and $X^{\mathcal{C}}$ commute and $-1$ otherwise.
The above expression allows one to write $\tr \rho\rho_{m}^{n-1}$ as
\begin{align}
    \tr \rho\rho_{m}^{n-1} &= \frac{\mathcal{Z}_{n,x}}{2^{(n-1)N}} \sum_{\{g_z^{(s)}\}}\mathcal{O}_D^{(n)} e^{-H_{n,z}},
\end{align}
where $\mathcal{O}_D^{(n)}$ denotes the product of sign functions in $n-1$ copies of $\rho_m$
\begin{align}
    \mathcal{O}_D^{(n)} = \sgn\big(g_z^{(1)}, X^\mathcal{C}\big).
\end{align}
Here, we have used the constraint $g_z^{(1)} = \prod_{s = 2}^{n} g_z^{(s)}$ for nonvanishing trace in the loop representation.
Using this expression, the $n$-th R\'enyi relative entropy takes the form
\begin{align}
    D^{(n)}(\rho||\rho_m) = \frac{1}{1-n}\log\langle \mathcal{O}_D^{(n)} \rangle\,.
\end{align}

Our next step is to express the observable $\langle \mathcal{O}_D^{(n)} \rangle$ in terms of the Ising spins.
In the spin model, the closed loop $g_z^{(1)}$ is identified with the domain wall of $\sigma_i^{(1)}$, and the Ising spins on two sides of $g_z^{(1)}$ anti-align. 
Thus, $\sigma_{i_l}^{(1)}$ and $\sigma_{i_r}^{(1)}$ on the two ends of the open string $\mathcal{C}$ is aligned if $g_z^{(1)}$ crosses $\mathcal{C}$ for even number of times and is anti-aligned otherwise.
The parity of the crossing is exactly measured by the sign function $\sgn(g_z^{(1)}, X^{\mathcal{C}})$.
Hence, the observable $\langle \mathcal{O}_D^{(n)} \rangle$ maps to the correlation function
\begin{align}
    \langle \mathcal{O}_D^{(n)} \rangle = \langle \sigma_{i_l}^{(1)} \sigma_{i_r}^{(1)} \rangle\,.
\end{align}

\subsection{Coherent information}
\label{sec:stat-mech_Ic}

We now develop a spin model description for the R\'enyi coherent information $I_c^{(n)}$ in Eq.~\eqref{eq:renyi_Ic}.
In the definition of coherent information, the system density matrix $\rho_Q$ is the error corrupted state $\rho$ in Sec.~\ref{sec:stat-mech model}, and its $n$-th moment is mapped to the partition function of the $(n-1)$-flavor Ising model on the torus.
Here, we show that the $n$-th moment of $\rho_{RQ}$ maps to the partition function of the same model with defects (domain walls) inserted along large loops on the torus.

First, we write down the initial state of the system $Q$ and the reference $R$.
We consider two reference qubits and two logical qubits in the ground state subspace, and maximally entangle them in a Bell state.
Let $s^{a=x,z}_l$ be the Pauli operator of two reference qubits, and $\bar{g}_{al}$ be the four logical operators
\begin{equation}
	\begin{tikzpicture}[baseline={(current bounding box.center)}]
		\small
		
		\draw  [>=stealth,->](-22pt,0pt)..controls (-22pt,-15pt) and (20pt,-15pt)..(22pt,0pt);
		\draw (22pt,0pt)..controls(20pt,18pt) and (-22pt,15pt)..(-22pt,0pt);
		
		\draw[white,line width=4] (-2pt,-3.3pt)..controls (-7pt,-4pt) and (-7pt,-19pt)..(-4pt,-20pt);
		
		\draw [black!50](0+12pt,0pt) arc (-60:-120:25pt);
		\draw [black!50](-3+12pt,-1.6pt) arc (50:130:15pt);
		
		\draw [line width=0.6,black!60]
		(0pt,20pt)..controls (20pt,20pt) and (34pt,12pt)..
		(34.5pt,0pt)..controls (34pt,-12pt) and (20pt,-20pt)..(0pt,-20pt)..
		controls (-20pt,-20pt) and (-34pt,-12pt)..(-34.5pt,0pt)..controls(-34pt,12pt) and (-20pt,20pt)..(0pt,20pt);
		
		\draw[>=stealth, mid arrow]  (-2pt,-3.3pt)..controls (-7pt,-4pt) and (-7pt,-19pt)..(-4pt,-20pt);
		\draw [dotted](-2pt,-3.3pt)..controls (1pt,-4pt) and (1pt,-19pt)..(-4pt,-20pt);
		
		\node at (15pt,16pt){\scriptsize $l_1$};
		\node at (-10pt,-24pt){\scriptsize $l_2$};
	\end{tikzpicture}\qquad
	\begin{aligned}
		\bar{g}_{zl} :=& \prod_{\ell \in l} Z_\ell\,, \\
		\bar{g}_{xl} :=& \prod_{\ell \in l^*} X_\ell\,,
	\end{aligned}\label{eq:toric_code_logical_opr}
\end{equation}
where $l = l_{1,2}$ and $l^* = l^*_{1,2}$ are on the original and dual lattice.
We consider the Bell state prepared as the $+1$ eigenstate of stabilizers $\bar{g}_{zl}s^z_l$ and $\bar{g}_{xl}s^x_l$, and write the initial density matrix for the system and reference as
\begin{align}
	\rho_{0,RQ} &= \prod_{l = l_1,l_2}\prod_{a = x,z}\frac{1 + \bar{g}_{al} s^a_l}{2}  \prod_s \frac{1 + A_s}{2} \prod_p \frac{1 + B_p}{2}\,. \label{eq:rhoRQ}
\end{align}

Here, we again work in the loop picture of $\rho_{0,RQ}$, and further factorize the density matrix into a product
\begin{equation}
    \rho_{0,RQ} =
    \frac{1}{2^{N+2}} \Gamma_{0,RQ}^x\Gamma_{0,RQ}^z\,,
\end{equation}
where $\Gamma_{0,RQ}^a$ is a summation of $a = x,z$ loops and takes the form
\begin{align}
    \Gamma_{0,RQ}^a = \sum_{g_a} g_a \prod_{l = l_1,l_2}\left(1+\bar{g}_{a,l}s^a_l\right).
\end{align}
In the error corrupted state $\rho_{RQ}$, the $X$ and $Z$ error channels act on $\Gamma_{0,RQ}^z$ and $\Gamma_{0,RQ}^x$, respectively, giving rise to $\rho_{RQ} = \Gamma^x_{RQ}\Gamma^z_{RQ}/2^{N+2}$ with
\begin{align}
    \Gamma^a_{RQ} = \sum_{g_a}\sum_{d_{al}=0,1} &e^{-\mu_a\left|\prod_{l=l_1,l_2}(\bar{g}_{al})^{d_{al}} g_a\right|} \nonumber \\
    &g_a  \prod_{l=l_1,l_2}(\bar{g}_{al}s^a_l)^{d_{al}},
\end{align}
where $d_{al}$ is a binary variable indicating whether the loop operator in the summation acts on the non-contractible loop $l$ of the torus.

Our next step is to write down the $n$-th moment of $\rho_{RQ}$ in the loop picture
\begin{align}
    \tr\rho_{RQ}^n = \frac{1}{2^{n(N+2)}} \tr\left( \left(\Gamma^x_{RQ}\right)^n\left(\Gamma^z_{RQ}\right)^n\right),
\end{align}
where each $\Gamma^{x(z)}_{RQ}$ is a sum over all possible $X(Z)$ loop operators with positive weights.
The product of loop operators from $n$ copies has a non-vanishing trace only if the product is identity.
This imposes the constraint on loop configurations and allows expressing the $n$-th moment as a sum of partition functions
\begin{align}
    \tr \rho_{RQ}^n = \frac{1}{2^{(n-1)(N+2)}}\prod_{a = x,z}\sum_{\mathbf{d}_{a1}\mathbf{d}_{a2}} \mathcal{Z}_{n,a}^{(\mathbf{d}_{a1},\mathbf{d}_{a2})},
\end{align}
where $\mathbf{d}_{al}$ with $l = 1,2$ is a $(n-1)$-component binary vector, the sum runs over all possible $\mathbf{d}_{al}$, and $\mathcal{Z}_{n,a}^{(\mathbf{d}_{a1},\mathbf{d}_{a2})} = \sum_{\{g_a^{(s)}\}} e^{- H_{n,a}^{(\mathbf{d}_{a1},\mathbf{d}_{a2})}}$ is the partition function with an effective Hamiltonian
\begin{equation}
\begin{aligned}
    H_{n,a}^{(\mathbf{d}_{1a},\mathbf{d}_{2a})} = \mu_a &\sum_{s = 1}^{n-1} \left| (\bar{g}_{a1}^{(s)})^{d_{a1,s}} (\bar{g}_{a2}^{(s)})^{d_{a2,s}} g_{a}^{(s)}\right| \\
    + \mu_a &\left|\prod_{s = 1}^{n-1} (\bar{g}_{a1}^{(s)})^{d_{a1,s}} (\bar{g}_{a2}^{(s)})^{d_{a2,s}} g_{a}^{(s)} \right|.\label{eq:large_loop_model}
\end{aligned}
\end{equation}
Here, $d_{al,s}$ denotes the $s$-th component of vector $\mathbf{d}_{al}$.

The loop model in Eq.~\eqref{eq:large_loop_model} can be identified with a classical spin model similar to Eq.~\eqref{eq:Heff_loop}. 
However, there is an important difference due to the presence of the homologically nontrivial loop $\bar{g}_{al}^{(s)}$.
Here, we interpret the homologically trivial loop $g_a^{(s)}$ as the Ising domain wall and $\bar{g}_{al}^{(s)}$ as a defect along the non-contractible loop.
The defect corresponds to flipping the sign of Ising coupling along a large loop.
Specifically, for $Z$ ($X$) loops on the original lattice, we introduce Ising spin on the plaquettes (vertices) such that
\begin{align}
    \left| (\bar{g}_{a1})^{d_{a1,s}}_\ell (\bar{g}_{a2})^{d_{a2,s}}_\ell g_{a,\ell}^{(s)} \right| = \frac{1 - (-1)^{\lambda_\ell^{(s)}} \sigma_i^{(s)}\sigma_j^{(s)}}{2},
\end{align}
where $i,j$ are connected by the link $\ell$, and $\lambda_\ell^{(s)} = | (\bar{g}_{a1})^{d_{a1,s}}_\ell (\bar{g}_{a2})^{d_{a2,s}}_\ell|$ is binary variable that denotes whether the defect goes through the link $\ell$.
This results in an effective Hamiltonian
\begin{equation}
\begin{aligned}
    H_{n,a}^{(\mathbf{d}_{1a},\mathbf{d}_{2a})} = & -J_a\sum_{\langle i,j\rangle} \sum_{s = 1}^{n-1} (-1)^{\lambda_\ell^{(s)}}\sigma_i^{(s)}\sigma_j^{(s)} \\ 
    & + \prod_{s = 1}^{n-1} (-1)^{\lambda_\ell^{(s)}} \sigma_i^{(s)} \sigma_j^{(s)}.
\end{aligned}
\end{equation}
Hence, $\mathcal{Z}_{n,a}^{(\mathbf{d}_{a1},\mathbf{d}_{a2})}$ becomes the partition function of the classical spin model with defects inserting along the non-contractible loops labeled by binary vectors $\mathbf{d}_{al}$.

The mapping developed above allows a spin model description for the $n$-th R\'enyi coherent information $I_c^{(n)}$.
The $n$-th moment of $\rho_{Q}$ is identified with the partition function with no defect, i.e. $\tr\rho_{Q}^n = \mathcal{Z}_{n,x}^{(\mathbf{0},\mathbf{0})}\mathcal{Z}_{n,z}^{(\mathbf{0},\mathbf{0})}/2^{(n-1)N}$.
Therefore, we have 
\begin{equation}
    I_c^{(n)} = \frac{1}{n-1} \sum_{a = x,z}\log \frac{\sum_{\mathbf{d}_{a1}\mathbf{d}_{a2}} \mathcal{Z}_{n,a}^{(\mathbf{d}_{a1},\mathbf{d}_{a2})}}{2^{n-1} \mathcal{Z}_{n,a}^{(\mathbf{0},\mathbf{0})}}.
\end{equation}
Thus, the R\'enyi coherent information is associated with the excess free energy of inserting defects along non-contractible loops
\begin{align}
    \Delta F_{n,a}^{(\mathbf{d}_{a1},\mathbf{d}_{a2})} := -\log \left( \mathcal{Z}_{n,a}^{(\mathbf{d}_{a1},\mathbf{d}_{a2})} / \mathcal{Z}_{n,a}^{(\mathbf{0},\mathbf{0})} \right).
\end{align}

\subsection{Entanglement negativity}
\label{sec:stat-mech_neg}

Here, we show that the R\'enyi negativity in the error-corrupted state maps to the excess free energy for aligning spins in the statistical mechanical model.
Specifically, we consider the case when only one type of error, e.g. bit-flip errors, is present.

The first step is to write down the partially transposed density matrix $\rho^{T_A}$.
We again work in the loop representation, where the error corrupted state is expressed as a sum of Pauli strings $g = g_xg_z$ in Eq.~\eqref{eq:rho_loop}.
The Pauli string $g$ is invariant under the partial transpose up to a sign factor $y_A(g) = (-1)^{N_Y}$ depending on the number $N_Y$ of Pauli-Y operators inside the subsystem $A$.
Hence, 
\begin{equation}
    \rho^{T_A} = \frac{1}{2^N} \sum_g y_A(g) e^{-\mu_x|g_x|-\mu_z|g_z|} g.
\end{equation}
Using the above expression, one can write down the $n$-th moment of $\rho^{T_A}$
\begin{equation}
    \tr \left(\rho^{T_A}\right)^{n} 
    = \frac{1}{2^{(n-1)N}} \sum_{\{g^{(s)}\}} \mathcal{O}_N^{(n)} e^{-H_{n,x} -H_{n,z}}.\label{eq:n_rhoTA}
\end{equation}
Here, similar to $\tr\rho^n$, the trace imposes a constraint on the loop operators $g^{(s)}$, and the summation runs over $g^{(s)}$ only in the first $n-1$ copies.
The sign factors collected from the partial transpose in each copy are combined in $\mathcal{O}_N^{(n)}$, 
\begin{align}
    \mathcal{O}_N^{(n)} = \left[\prod_{s = 1}^{n-1} y_A\Big(g^{(s)}\Big)\right] y_A\Big(\prod_{s=1}^{n-1}g^{(s)}\Big). \label{eq:ON_yA}
\end{align}
%
%
\eqnref{eq:n_rhoTA} allows expressing the $2n$-th R\'enyi negativity in terms of the expectation value of $\mathcal{O}_{2n}$:
\begin{align}
    \calE^{(2n)}_{A} = \frac{1}{2-2n}\log \left\langle \mathcal{O}_N^{(2n)} \right\rangle.
\end{align}

Yet, analyzing the number of Pauli-Y operators in Eq.~\eqref{eq:ON_yA} is a formidable task.
Moreover, the observable $\mathcal{O}_N^{(n)}$ derived from the partial transpose should be a basis-independent quantity. 
Indeed, one can express $O_N^{(n)}$ in terms of loop configurations
\begin{align}
    \mathcal{O}_N^{(n)} &= \prod_{r = 1}^{n-2} \sgn_A\Big(\prod_{s = 1}^r g^{(s)},g^{(r+1)}\Big) \nonumber \\
    &= \prod_{r = 2}^{n-1} \prod_{s = 1}^{r-1}\sgn_A\big(g^{(s)}, g^{(r)}\big). \label{eq:ON_sgn1}
\end{align}
Here, we use the property 
\begin{equation}
y_A(g)y_A(h) = y_A(gh) \sgn_A(g,h),
\end{equation}
where the sign function $\sgn_A(g,h) = \pm 1$ depending on the commutation relation between the support of Pauli string $g$ and $h$ on subsystem $A$:
\begin{align}
    \sgn_A(g,h) = \left\{ \begin{array}{cc}
    1 & [g_A, \; h_A] = 0 \\
    -1 & \{g_A, \; h_A\} = 0
    \end{array}\right. .
\end{align}
In the second equality of Eq.~\eqref{eq:ON_sgn1}, we use the property of sign function \begin{equation}
\sgn_A(g_1g_2, g_3) = \sgn_A(g_1,g_3) \sgn_A(g_2,g_3).
\end{equation}

In the Toric code, the operator $g$ further factorizes into $g = g_x g_z$, where $g_x, g_z$ are closed loop operators of Pauli $X$ and $Z$, respectively. 
The sign function between two such loop operators $g$ and $h$ reduces to
\begin{align}
    \sgn_A(g,h) = \sgn_A(g_x, h_z) \sgn_A(g_z, h_x).
\end{align}
We then arrive at 
\begin{align}
    \mathcal{O}_N^{(n)} = \prod_{s,r = 1, s \neq r}^{n-1} \sgn_A\big(g_x^{(s)}, g_z^{(r)}\big).\label{eq:ON_sgn}
\end{align}

To develop an analytic understanding of the observable $\mathcal{O}_N^{(n)}$ and how it detects the ferromagnetic transition, we first consider the situation when only $X$ or $Z$ error is present.
In this case, we show that $\log\langle\mathcal{O}_N^{(n)}\rangle$ exactly maps to the excess free energy of spin pinning and sharply distinguish the two phases.
After that, we discuss the general situation when both types of error are present.

We here consider the case when only $X$ errors are present, namely $p_z = 0$ and $\mu_x = 0$.
The vanishing $X$-loop tension indicates that $H_{n,x}$ is in the paramagnetic phase, and the domain walls $g_x$ of arbitrary sizes occur with the same probability.
Thus, we can perform an exact summation over all possible $g_x$ and obtain
\begin{equation}
    \tr (\rho^{T_A})^n = \frac{1}{2^{(n-1)N}}  \sum_{\{g_z^{(s)}\}} \mathcal{O}^{(n)}_{N,z} e^{-\mu_z H_{n,z}},
\end{equation}
where $\mathcal{O}^{(n)}_{N,z} = \sum_{\{g_x^{(s)}\}} \mathcal{O}^{(n)}_N$.
The summation in $\mathcal{O}^{(n)}_{N,z}$ is non-vanishing only if the sign functions in Eq.~\eqref{eq:ON_sgn} for different $g_x^{(s)}$ interfere constructively.
This yields a constraint on the $g_{z}^{(s)}$
\begin{align}
    \mathcal{O}^{(n)}_{N,z} = \prod_{r = 1}^{n-1} N_{g_x} \delta_{h^{(r)}(A)}
\end{align}
where $h^{(r)} = \prod_{s = 1, s \neq r}^{n-1} g_z^{(s)}$, the Kronecker delta function $\delta_{h^{(r)}(A)}$ takes the value unity only if the support of $h^{(r)}$ on subsystem $A$ is a closed loop and equals zero otherwise, and $N_{g_x}$ is an unimportant prefactor that denotes the number of possible $g_x$ in each copy.
The $n-1$ delta function constraints are independent for odd $n$, whereas for even $n$ they give rise to only $n - 2$ independent constraints as $\prod_{r = 1}^{n-1} h^{(r)} = I$.

The constraint requires $h^{(r)}$ not to go through the boundary of subsystem $A$.
In the statistical mechanical model of Ising spins, this corresponds to no domain wall going through the boundary of $A$, namely forcing $|\partial A|$ boundary spins aligning in the same direction (see Fig.~\ref{fig:neg_constraint}).
Thus, the negativity is associated with the excess free energy for aligning spins
\begin{align}
\calE_{A}^{(2n)} = \frac{1}{2n-2}(F_A^{(2n)}-F_0^{(2n)}) := \frac{\Delta F_A^{(2n)}}{2n-2},
\end{align}
where $F_0^{(2n)} := -\log \calZ_{2n,x}\calZ_{2n,z}$ and $F_A^{(2n)}$ are the free energy without and with constraints, respectively.
Since we have in total $2n-2$ constraints, $\calE_A^{(2n)} = \Delta F_A$ with $\Delta F_A$ being the excess free energy for aligning one species of Ising spins.

\section{$\mathbb{Z}_N$ Toric code}
\label{app:Zn toric code}

So far, we only focus on the $\bbZ_2$ Toric code with incoherent errors. 
It is natural to inquire whether our methods are still applicable to $\bbZ_N$ Toric code and whether the results change. We provide a brief discussion on the $\bbZ_3$ Toric code in this subsection. 
We will use similar symbols to denote the basic operators and stabilizers, although their meanings are different from those in the $\bbZ_2$ case.

Let us first specify the Hamiltonian and the error models.
Consider an $L\times L$ square lattice with periodic boundary conditions. The physical qutrits live on the edges of the lattice.
We introduce the clock and shift operators
\begin{equation}
\begin{gathered}
XZ = w ZX\,,\quad w = e^{2\pi i/3}\,, \\
Z = \begin{pmatrix} 1 \\ & w \\ && w^2 \end{pmatrix},\quad
X = \begin{pmatrix} & 1 \\ & & 1 \\ 1 &&  \end{pmatrix}\,.
\end{gathered}
\end{equation}
In and only in this subsection, $X$ and $Z$ refer to the clock and shift, respectively.
The code subspace is given by the ground state subspace of the Hamiltonian
\begin{equation}
    H_{\bbZ_3} = -\sum_s A_s - \sum_p B_p
\end{equation}
where $A_s$ and $B_p$ are mutually commuting projectors associated with vertices and plaquettes, e.g.,
\begin{equation}
\begin{tikzpicture}[scale=0.7, baseline=15]
\foreach \y in {0,1,2}{
	\draw (-0.2,\y) -- (2.2,\y);
}
\foreach \x in {0,1,2}{
	\draw (\x,-0.2) -- (\x,2.2);
}
\filldraw[black] (1,1) circle (0.1) node[red,below right]{\scriptsize $s$};
\node[red] at (0.5,1.5) {\scriptsize$p$};
\draw[mid arrow,line width=1.5] (1,1) -- node[right]{\scriptsize1} (1,2);
\draw[mid arrow,line width=1.5] (0,2) -- node[above]{\scriptsize2} (1,2);
\draw[mid arrow,line width=1.5] (0,1) -- node[left]{\scriptsize3} (0,2);
\draw[mid arrow,line width=1.5] (0,1) -- node[below]{\scriptsize4} (1,1);
\draw[mid arrow,line width=1.5] (1,0) node[above right]{\scriptsize5} -- (1,1);
\draw[mid arrow,line width=1.5] (1,1) -- (2,1) node[below left]{\scriptsize6};
\end{tikzpicture}\quad
\begin{aligned}
A_s =& \frac{1}{3}\sum_{n=0}^2 \big(X_4 X_5 X_1^{-1} X_6^{-1} \big)^n \\
B_p =& \frac{1}{3}\sum_{n=0}^2 \big(Z_4 Z_1 Z_2^{-1} Z_3^{-1} \big)^n
\end{aligned}
\end{equation}
One can verify that $A_s^2 = A_s$, $B_p^2 = B_p$.
The ground state $\ket{\Psi}$ satisfies $A_s \ket{\Psi} = B_p \ket{\Psi} = \ket{\Psi}$, and the violation of $A_s$ and $B_p$ will be refered to as $e$ (and its anti-particle $\bar{e}$) and $m$ (and its anti-particle $\bar{m}$) anyons, respectively.
For simplicity, we only consider the following incoherent error
\begin{equation}
\begin{aligned}
\calN_{X,i}[\rho] = (1- & p_1-p_2) \rho \\
+ & p_1 Z_i \rho Z_i^\dag + p_2 Z_i^2 \rho Z_i^{2,\dag}\,,
\end{aligned}
\end{equation}
which creates a pair of $e$ anyons in two different ways with probabilities $p_1$ and $p_2$.
In the following, we will first assume $p_1=p_2=p$ and comment on what could change without this assumption.

To compute the three diagnostics, one can still work in the loop picture and map the $n$-th momentum of the error-corrupted state to a partition function of a classical spin model that involves $n$-flavor 3-state Potts spins. 
As the error rate increases, the spin model undergoes a paramagnet-to-ferromagnet transition.
The three diagnostics are mapped to the corresponding observables in a similar fashion as what we have shown in the $\bbZ_2$ case.
Therefore, they should undergo a transition simultaneously and yield a consistent characterization of the error-induced phase.

When $p_1\neq p_2$, the spin models obtained in the loop picture contain complex phases and do not admit a statistical mechanical interpretation. Technically, it brings sign problems to the Monte Carlo simulation.
It is unclear whether the three diagnostics still exhibit transition simultaneously, which may be an interesting question for future study.

\bibliography{ref.bib}
\end{document}